\documentclass[nohyperref]{article}

\usepackage{microtype}
\usepackage{graphicx}
\usepackage{caption}
\usepackage{subcaption}
\usepackage{balance}
\usepackage{booktabs} %

\usepackage{hyperref}

\usepackage[accepted]{icml2022}  %

\usepackage{amsmath}
\usepackage{amssymb}
\usepackage{mathtools}
\usepackage{amsthm}
\usepackage[capitalize,noabbrev]{cleveref}

\theoremstyle{plain}
\newtheorem{theorem}{Theorem}[section]
\newtheorem{proposition}[theorem]{Proposition}
\newtheorem{lemma}[theorem]{Lemma}
\newtheorem{corollary}[theorem]{Corollary}

\theoremstyle{definition}
\newtheorem{definition}[theorem]{Definition}

\theoremstyle{remark}

\icmltitlerunning{Faster Privacy Accounting via Evolving Discretization}

\def\tdotoggle{1}
\usepackage{xspace}
\usepackage{enumitem}
\usepackage{thmtools,thm-restate}

\setlist[itemize]{leftmargin=*,label=$\triangleright$,itemsep=0pt}

\newcommand{\polylog}{\mathrm{polylog}}

\definecolor{pw}{HTML}{7977B8}
\definecolor{og}{HTML}{3C8031}
\definecolor{maroon}{HTML}{AF3235}
\definecolor{yo}{HTML}{FAA21A}
\definecolor{mybrick}{RGB}{180,14,15}
\definecolor{Gred}{RGB}{219, 50, 54}
\definecolor{Ggreen}{RGB}{60, 186, 84}
\definecolor{Gblue}{RGB}{72, 133, 237}
\definecolor{Gyellow}{RGB}{247, 178, 16}
\definecolor{ToCgreen}{RGB}{0, 128, 0}
\definecolor{myGold}{RGB}{231,141,20}
\definecolor{myBlue}{rgb}{0.19,0.41,.65}
\definecolor{myPurple}{RGB}{175,0,124}

\DeclareMathOperator{\Ex}{\mathbb{E}}

\newcommand{\what}[1]{\widehat{#1}}
\newcommand{\wtilde}[1]{\widetilde{#1}}

\renewcommand{\mod}[1]{\ (\mathrm{mod}\ #1)}
\newcommand{\dTV}{d_{\mathrm{TV}}}

\def\eps{\varepsilon}

\newcommand{\set}[1]{\left \{ #1 \right \}}

\newcommand{\inabs}[1]{\left | #1 \right |}
\newcommand{\inparen}[1]{\left ( #1 \right )}
\newcommand{\insquare}[1]{\left [ #1 \right ]}

\newcommand{\infork}[1]{\left \{ \begin{matrix} #1 \end{matrix} \right .}
\newcommand{\inmat}[1]{\begin{matrix} #1 \end{matrix}}

\newcommand{\floor}[1]{\left \lfloor #1 \right \rfloor}

\newcommand{\bbR}{\mathbb{R}}

\newcommand{\bbZ}{\mathbb{Z}}

\newcommand{\calM}{\mathcal{M}}
\newcommand{\calN}{\mathcal{N}}

\usepackage{inconsolata}

\newcommand{\PDF}{\mathsf{PDF}}
\newcommand{\CDF}{\mathsf{CDF}}
\newcommand{\TwoStageSelfComposePRV}{\mathsf{TwoStageSelfComposePRV}}
\newcommand{\TwoStageComposePRV}{\mathsf{TwoStageComposePRV}}
\newcommand{\RecursiveSelfComposePRV}{\mathsf{RecursiveSelfComposePRV}}
\newcommand{\DiscretizeRV}{\mathsf{DiscretizeRV}}

\newcommand{\epsupper}{\eps_{\mathrm{up}}}
\newcommand{\epserror}{\eps_{\mathrm{err}}}
\newcommand{\deltaerror}{\delta_{\mathrm{err}}}

\begin{document}

\twocolumn[
\icmltitle{Faster Privacy Accounting via Evolving Discretization}

\icmlsetsymbol{equal}{*}

\begin{icmlauthorlist}
\icmlauthor{Badih Ghazi}{google}
\icmlauthor{Pritish Kamath}{google}
\icmlauthor{Ravi Kumar}{google}
\icmlauthor{Pasin Manurangsi}{google}
\end{icmlauthorlist}

\icmlaffiliation{google}{Google Research, USA}

\icmlcorrespondingauthor{Pritish Kamath}{pritish@alum.mit.edu}
\icmlcorrespondingauthor{Pasin Manurangsi}{pasin@google.com}

\icmlkeywords{Machine Learning, ICML}

\vskip 0.3in
]

\printAffiliationsAndNotice{\icmlEqualContribution} %

\begin{abstract}
We introduce a new algorithm for numerical composition of privacy random variables, useful for computing the accurate differential privacy parameters for composition of mechanisms.
Our algorithm achieves a running time and memory usage of $\mathrm{polylog}(k)$ for the task of self-composing a mechanism, from a broad class of mechanisms, $k$ times; this class, e.g., includes the sub-sampled Gaussian mechanism, that appears in the analysis of differentially private stochastic gradient descent.
By comparison, recent work by \citet{gopi2021numerical} has obtained a running time of $\widetilde{O}(\sqrt{k})$ for the same task.
Our approach extends to the case of composing $k$ different mechanisms in the same class, improving upon their running time and memory usage from $\widetilde{O}(k^{1.5})$ to $\widetilde{O}(k)$.
\end{abstract}

\section{Introduction}

Differential privacy (DP)~\cite{dwork06calibrating,dwork2006our} has become the preferred notion of privacy in both academia and the industry.  Fueled by the increased awareness and  demand for privacy, several systems that use DP mechanisms to guard users' privacy have been deployed in the industry~\cite{erlingsson2014rappor,CNET2014Google, greenberg2016apple,dp2017learning, ding2017collecting, LinkedINDP1, LinkedInDP2}, and the US Census~\cite{abowd2018us}.  Besides the large volume of data, many of these systems offer real-time private data analytic and inference capabilities, which entail strict computational efficiency requirements on the underlying DP operations. 

We recall the definition of DP~\cite{dwork06calibrating,dwork2006our}:
\begin{definition}[DP]
    Let $\eps > 0$ and $\delta \in [0, 1]$.  
    A randomized algorithm $\calM : \mathcal{X}^n \to \mathcal{Y}$ is \emph{$(\eps,\delta)$-DP} if, for all $x,x' \in \mathcal{X}^n$ differing on a single index and all outputs $S \subseteq \mathcal{Y}$, we have $\Pr[\calM(x)\in S] \le e^\eps \cdot \Pr[\calM(x')\in S]+\delta$.
\end{definition}
The DP guarantees of a mechanism are captured by the privacy \emph{parameters} $\eps$ and $\delta$; the smaller these parameters, the more private the mechanism.  Often a mechanism is simultaneously DP for multiple privacy parameters; this is captured by studying the \emph{privacy loss} of a mechanism $\calM$---the curve $\delta_{\calM}(\cdot)$ for which the mechanism is $(\eps, \delta_{\calM}(\eps))$-DP.

A fundamental mathematical property satisfied by DP is \emph{composition}, which prescribes the privacy guarantees of results from executing multiple DP mechanisms.  In \emph{basic} composition~\cite{dwork2006our}, a mechanism that returns the results of executing an $(\eps_1, \delta_1)$-DP mechanism and an $(\eps_2, \delta_2)$-DP mechanism is $(\eps_1+\eps_2, \delta_1+\delta_2)$-DP.  Unfortunately, this bound becomes weak when composing a large number of mechanisms.  The \emph{advanced} composition~\cite{dwork2010boosting} offers stronger bounds: roughly speaking, composing $k$ mechanisms that are each $(\eps,\delta)$-DP yields a mechanism whose privacy parameters are of the order of $\sqrt{k} \eps$ and $k \delta$---clearly more desirable than the basic composition.  In fact, obtaining tight composition bounds has been an active research topic. \citet{kairouz2015composition} showed how to obtain the exact privacy parameters of self-composition (mechanism composed with itself), while~\citet{murtagh2016complexity} showed that the corresponding problem for the more general case is \#P-complete.  

\paragraph{Privacy Loss Distribution (PLD).}
Tighter bounds for privacy parameters that go beyond advanced composition are possible if the privacy loss of the mechanism is taken into account.  
%
\citet{meiser2018tight, sommer2019privacy} initiated the study of numerical methods for accurately estimating the privacy parameters, using the \emph{privacy loss distribution} (PLD) of a mechanism. The PLD is the probability mass function of the so-called \emph{privacy loss random variable} (PRV) for discrete mechanisms, and its density function for continuous mechanisms (see Section~\ref{sec:prelims} for formal definitions).  PLDs have two nice properties: (i) tight privacy parameters can be computed from the PLD of a mechanism and (ii) the PLD of a composed mechanism is the convolution of the individual PLDs.  Property (ii) makes PLD amenable to  FFT-based methods.

\citet{koskela2020computing} exploited this property to speed up the computation of the PLD of the composition.  
An important step to retain efficiency using PLDs is approximating the distribution so that it has finite support; this is especially needed in the case when the PLD is continuous or has a large support.  PLD implementations have been at the heart of many DP systems and open-source libraries including \cite{DPBayes, GoogleDP, MicrosoftDP}.  To enable scale and support latency considerations, the PLD composition has to be as efficient as possible.  This is the primary focus of our paper.

Our starting point is the work of~\citet{koskela2020computing,koskela2021tight,koskela2021computing}, who derived
explicit error bounds for the approximation obtained by the FFT-based algorithm. The running time of this algorithm for $k$-fold self-composition of a mechanism $\calM$ that is 
$(\eps_0, 0)$-DP 
is\footnote{For any positive $T$, we denote by $\wtilde{O}(T)$ any quantity of the form $O(T \cdot \polylog(T))$.} $\wtilde{O}\inparen{\frac{k^{2} \eps_0}{\deltaerror}}$. When $\eps_0 = \frac{1}{\sqrt{k \cdot \log(1/\deltaerror)}}$, this running time is $\wtilde{O}\inparen{\frac{k^{1.5}}{\deltaerror}}$, where $\deltaerror$ is the additive error in $\delta$.  \citet{gopi2021numerical} improved this bound to obtain an algorithm with running time of
\[\textstyle\wtilde{O}\inparen{\frac{k^{0.5} \sqrt{\log \frac{1}{\deltaerror}}}{\epserror}}\,,\]
where $\epserror$ is the additive error in $\eps$. When composing $k$ different mechanisms, their running time is 
\[\textstyle\wtilde{O}\inparen{\frac{k^{1.5} \sqrt{\log \frac{1}{\deltaerror}}}{\epserror}}\,.\]

\paragraph{Our Contributions.}
We design and study new algorithms for $k$-fold numerical composition of PLDs.  Specifically, for reasonable choice of mechanisms, we
\begin{itemize}[nosep]
\item obtain (\cref{sec:two-stage-compose}) a two-stage algorithm for self-composing PLDs, with running time
\[\wtilde{O}\inparen{\frac{k^{0.25}\sqrt{\log \frac{1}{\deltaerror}}}{\epserror}}\,.\]
\item provide (\cref{sec:experiments}) an experimental evaluation  of the two-stage algorithm, comparing its runtime to that of the algorithm of~\citet{gopi2021numerical}. We find that the speedup obtained by our algorithm improves with $k$.
\item extend (\cref{sec:recursive-compose}) the two-stage algorithm to a recursive multi-stage algorithm with a running time of
\[\wtilde{O}\inparen{\frac{\polylog(k) \sqrt{\log \frac{1}{\deltaerror}}}{\epserror}}\,.\]
\end{itemize}
Both the two-stage and multi-stage algorithms extend to the case of composing $k$ different mechanisms. In each case, the running time increases by a multiplicative $O(k)$ factor. Note that this factor is inevitable since the input ``size'' is $k$---indeed, the algorithm needs to read the $k$ input PLDs. We defer the details of this extension to \cref{apx:two-stage-extensions}.

\paragraph{Algorithm Overview.}

The main technique underlying our algorithms is the evolution of the discretization and truncation intervals of the approximations of the PRV with the number of compositions.  To describe our approach, we first present a high-level description of the algorithm of~\citet{gopi2021numerical}.  Their algorithm discretizes an $O(1)$-length interval into bucket intervals each with mesh size $\approx \frac{1}{k^{0.5}}$, thus leading to a total of $O(k^{0.5})$ buckets and a running time of $\wtilde{O}(k^{0.5})$ for the FFT convolution algorithm. Both these aspects of their algorithm are in some sense necessary: a truncation interval of length $\ll O(1)$ would lose significant information about the $k$-fold composition PRV, whereas a discretization interval of length $\gg \frac{1}{k^{0.5}}$ would lose significant information about the original PRV; so relaxing either would lead to large approximation error.

The key insight in our work is that it is possible to avoid having both these aspects {\em simultaneously}. In particular, in our two-stage algorithm, the first stage performs a $k^{0.5}$-fold composition using an interval of length $\approx \frac{1}{k^{0.25}}$ discretized into bucket intervals with mesh size $\approx \frac{1}{k^{0.5}}$, followed by another $k^{0.5}$-fold composition using an interval of length $O(1)$ discretized into bucket intervals with mesh size $\approx \frac{1}{k^{0.25}}$. Thus, in each stage the number of discretization buckets is $\wtilde{O}(k^{0.25})$ leading to a final running time of $\wtilde{O}(k^{0.25})$ for performing two FFT convolutions.

The recursive multi-stage algorithm extends this idea to $O(\log k)$ stages, each performed with an increasing discretization interval and truncation interval, ensuring that the number of discretization buckets at each step is $O(\polylog(k))$.

\paragraph{Experimental Evaluation.}
We implement our two-stage algorithm and compare it to baselines from the literature. We consider the sub-sampled Gaussian mechanism, which is a fundamental primitive in private machine learning and constitutes a core primitive of training algorithms that use differentially private stochastic gradient descent (DP-SGD) (see, e.g., \cite{abadi2016deep}). For $2^{16}$ compositions and a standard deviation of $\approx 226.86$ and with subsampling probability of $0.2$, we obtain a speedup of $2.66 \times$ compared to the state-of-the-art. We also consider compositions of the widely-used Laplace mechanism. For $2^{16}$ compositions, and a scale parameter of $\approx 1133.84$ for the Laplace distribution, we achieve a speedup of $2.3 \times$. The parameters were chosen such that the composed mechanism satisfies $(\eps=1.0, \delta=10^{-6})$-DP. See \Cref{sec:experiments} for more details.

\paragraph{Related Work.}
In addition to Moments Accountant~\cite{abadi2016deep} and R\'{e}nyi DP~\cite{mironov2017renyi} (which were originally used to bound the privacy loss in DP deep learning), several other tools can also be used to upper-bound the privacy parameters of composed mechanisms, including concentrated DP~\cite{dwork2016concentrated, bun2016concentrated}, its truncated variant~\cite{bun2018composable}, and Gaussian DP \cite{dong2019gaussian, bu2020deep}. However, none of these methods is tight; furthermore, none of them allows a high-accuracy estimation of the privacy parameters.  In fact, some of them require that the moments of the PRV are bounded; the PLD composition approach does not have such restrictions and hence is applicable to mechanisms such as DP-SGD-JL~\cite{bu2021fast}.

In a recent work,~\citet{ours-pets} proposed a different discretization procedure based on whether we want the discretized PLD to be ``optimistic'' or ``pessimistic''. They do not analyze the error bounds explicitly but it can be seen that their running time is $\tilde{O}(k)$, which is slower than both our algorithm and that of~\citet{gopi2021numerical}.

Another recent work~\cite{zhu21optimal} developed a rigorous notion of ``worst-case'' PLD for mechanisms, under the name \emph{dominating PLDs}. Our algorithms can be used for compositions of dominating PLDs; indeed, our experimental results for Laplace and (subsampled) Gaussian mechanisms are already doing this implicitly. %
Furthermore,~\citet{zhu21optimal} propose a different method for computing tight DP composition bounds. However, their algorithm requires an analytical expression for the characteristic function of the PLDs.  This may not exist, e.g., we are unaware of such an analytical expression for subsampled Gaussian mechanisms.

\section{Preliminaries}\label{sec:prelims}
Let $\bbZ_{> 0}$ denote the set of all positive integers, $\mathbb{R}_{\geq 0}$ the set of all non-negative real numbers, and let $\overline{\mathbb{R}} = \mathbb{R} \cup \{- \infty, +\infty\}$. For any two random variables $X$ and $Y$, we denote by $\dTV(X, Y)$ their \emph{total variation distance}.

\subsection{Privacy Loss Random Variables}
We will use the following privacy definitions and tools that appeared in previous works on PLDs \cite{sommer2019privacy,koskela2020computing, gopi2021numerical}. 

For any mechanism $\mathcal{M}$, and any $\eps \in \mathbb{R}_{\geq 0}$, we denote by $\delta_{\mathcal{M}}(\eps)$ the smallest value $\delta$ such that $\mathcal{M}$ is $(\eps, \delta)$-DP. The resulting function $\delta_{\mathcal{M}}(\cdot)$ is said to be the \emph{privacy curve} of the mechanism $\mathcal{M}$. A closely related notion is the \emph{privacy curve} $\delta(X || Y): \bbR_{\geq 0} \to [0, 1]$ between two random variables $X, Y$, and which is defined, for any $\eps \in \mathbb{R}_{\ge 0}$ as
\begin{equation*}
    \delta(X || Y)(\eps) = \sup_{S \subset \Omega} \Pr[Y \in S] - e^{\eps} \Pr[X \in S],
\end{equation*}
where $\Omega$ is the support of $X$ and $Y$.
The \emph{privacy loss random variables} associated with a privacy curve $\delta_{\cal M}$ are random variables $(X, Y)$ such that $\delta_{\cal M}$ is the same curve as $\delta(X || Y)$, and that satisfy certain additional properties (which make them unique). While PRVs have been defined earlier in \citet{dwork2016concentrated,bun2016concentrated}, we use the definition of \citet{gopi2021numerical}:
\begin{definition}[PRV]
Given a privacy curve $\delta_{\cal M}: \mathbb{R}_{\geq 0} \to [0,1]$, we say that random variables $(X, Y)$ are   \emph{privacy loss random variables (PRVs)} for $\delta_{\cal M}$, if (i) $X$ and $Y$ are supported on $\overline{\mathbb{R}}$, (ii) $\delta(X||Y) = \delta_{\cal M}$, (iii) $Y(t) = e^t X(t)$ for every $t \in \mathbb{R}$, and (iv) $Y(-\infty) = X(\infty) = 0$, where $X(t)$ and $Y(t)$ denote the PDFs of $X$ and $Y$, respectively.
\end{definition}

\begin{theorem}[\citet{gopi2021numerical}]
    Let $\delta$ be a privacy curve that is identical to $\delta(P || Q)$ for some random variables $P$ and $Q$. Then, the PRVs $(X,Y)$ for $\delta$ are given by
    \[ \textstyle
        X = \log\inparen{\frac{Q(w)}{P(w)}} \text{ where } \omega \sim P,
    \]
    and
    \[ \textstyle
        Y = \log\inparen{\frac{Q(w)}{P(w)}} \text{ where } \omega \sim Q.
    \]
Moreover, we define $\delta_Y(\eps) := \Ex_Y [(1 - e^{\eps - Y})_+]$ for every $\eps \in \mathbb{R}$ and define $\eps_Y(\delta)$ as the smallest $\eps$ such that $\delta_Y(\eps) \le \delta$.
\end{theorem}

Note that $\delta_Y(\eps)$ is well defined even for $Y$ that is \emph{not} a privacy loss random variable.

If $\delta_1$ is a privacy curve identical to $\delta(X_1 || Y_1)$ and $\delta_2$ is a privacy curve identical to $\delta(X_2 || Y_2)$, then the composition $\delta_1 \otimes \delta_2$ is defined as the privacy curve $\delta((X_1, X_2) || (Y_1, Y_2))$, where $X_1$ is independent of $X_2$, and $Y_1$ is independent of $Y_2$. A crucial property is that composition of privacy curves corresponds to addition of PRVs:
\begin{theorem}[\citet{dwork2016concentrated}]
Let $\delta_1$ and $\delta_2$ be privacy curves with PRVs $(X_1, Y_1)$ and $(X_2, Y_2)$ respectively. Then, the PRVs for the privacy curve $\delta_1 \otimes \delta_2$ are given by $(X_1 + X_2, Y_1 + Y_2).$
\end{theorem}

We interchangeably use the same letter to denote both a random variable and its corresponding probability distribution. For any two distributions $Y_1, Y_2$, we use $Y_1 \oplus Y_2$ to denote its \emph{convolution} (same as the random variable $Y_1 + Y_2$).  We use $Y^{\oplus k}$ to denote the $k$-fold convolution of $Y$ with itself.

Finally, we use the following tail bounds for PRVs.
\begin{lemma}[\citet{gopi2021numerical}]\label{lem:tail-bounds}
	For all PRV $Y$, $\eps \ge 0$ and $\alpha > 0$, it holds that\vspace{-2mm}
	\begin{align*}
	\textstyle \Pr[|Y| \ge \eps + \alpha]
	&\textstyle~\le~ \delta_Y(\eps) \cdot \frac{(1 + e^{-\eps-\alpha})}{1 - e^{-\alpha}}\\
	&\textstyle~\le~ \frac{4}{\alpha} \delta_Y(\eps) \quad \text{if } \alpha < 1.
	\end{align*}
\end{lemma}

\subsection{Coupling Approximation}

To describe and analyze our algorithm, we use the coupling approximation tool used by~\citet{gopi2021numerical}.  They showed that, in order to provide an approximation guarantee on the privacy loss curve, it suffices to approximate a PRV according to the following coupling notion of approximation:
\begin{definition}[Coupling Approximation]
\label{def:coupling}
For two random variables $Y_1, Y_2$, we write $|Y_1 - Y_2| \le_{\eta} h$ if there exists a \emph{coupling} between them such that $\Pr[|Y_1 - Y_2| > h] \le \eta$.
\end{definition}
We use the following properties of coupling approximation shown by \citet{gopi2021numerical}. We provide the proofs in \Cref{apx:gopi-proofs} for completeness.
\begin{lemma}[Properties of Coupling Approximation]
\label{lem:coupling-properties}
Let $X, Y, Z$ be random variables.
\begin{enumerate}[leftmargin=*,nosep]
\item \label{item:coupling-to-delta}
If $|X - Y| \leq_{\deltaerror} \epserror$, then for all $\eps \in \mathbb{R}_{\geq 0}$, 
\[ \delta_{Y}(\eps + \epserror) - \deltaerror
~\le~ \delta_X(\eps) ~\le~
\delta_{Y}(\eps - \epserror) + \deltaerror. \]
\item \label{item:coupling-triangle}
If $|X - Y| \le_{\eta_1} h_1$ and $|Y - Z| \le_{\eta_2} h_2$, then
$|X - Z| \le_{\eta_1 + \eta_2} h_1 + h_2$ (``triangle inequality'').
\item \label{item:coupling-composition}
If $\dTV(X, Y) \le \eta$, then $|X - Y| \le_{\eta} 0$; furthermore, for all $k \in \mathbb{Z}_{> 0}$, it holds that $|X^{\oplus k} - Y^{\oplus k}| \le_{k\eta} 0$.
\item \label{item:coupling-concentration}
If $|X - Y - \mu| \le_0 h$ (for any $\mu$) and $\Ex[X] = \Ex[Y]$, then for all $\eta > 0$ and  $k\in \mathbb{Z}_{> 0}$,
\[	\textstyle
\inabs{X^{\oplus k} - Y^{\oplus k}} ~\le_{\eta}~ h \sqrt{2k \log \frac{2}{\eta}}.
\]
\end{enumerate}
\end{lemma}

\subsection{Discretization Procedure}

We adapt the discretization procedure from \cite{gopi2021numerical}.  The only difference is that we assume the support of the input distribution is already in the specified range as opposed to being truncated as part of the algorithm.  A complete description of the procedure is given in Algorithm~\ref{alg:discretize-prv}. 

\begin{algorithm}[ht]
	\caption{$\DiscretizeRV$ \citep[adapted from][]{gopi2021numerical}}
	\label{alg:discretize-prv}
	\begin{algorithmic}
		\STATE {\bfseries Input:} $\CDF_Y(\cdot)$ of a RV $Y$, mesh size $h$, truncation parameter $L \in h \cdot (\frac12 + \bbZ_{> 0})$.
		\STATE {\bfseries Constraint:} Support of $Y$ is contained in $(-L, L]$.
		\STATE {\bfseries Output:} $\PDF$ of an approximation $\wtilde{Y}$ supported on $\mu + (h\bbZ \cap [-L, L])$ for some $\mu \in [-\frac{h}{2}, \frac{h}{2}]$.
		\STATE
		\STATE $n \gets \frac{L - \frac{h}{2}}{h}$
		\FOR{$i=-n$ {\bfseries to} $n$}
		\STATE $q_i \gets \CDF_Y(ih + h/2) - \CDF_Y(ih - h/2)$
		\ENDFOR
		\STATE $q \gets q / (\sum_{i=-n}^n q_i)$ \hfill \texttt{\small $\triangleright$ Normalize $q$}
		\STATE $\mu \gets \Ex[Y] - \sum_{i=-n}^n ih \cdot q_i$
		\STATE $\wtilde{Y} \gets \infork{ih + \mu & \text{w.p. } q_i \quad \text{ for } -n \le i \le n}$
		\RETURN $\wtilde{Y}$
	\end{algorithmic}
\end{algorithm}

The procedure can be shown to satisfy the following key property.
\begin{proposition} \label{claim:discretize-prv-prop}
	For any random variable $Y$ supported in $(-L, L]$, the output $\wtilde{Y}$ of $\DiscretizeRV$ with mesh size $h$ and truncation parameter $L$ satisfies $\Ex[Y] = \Ex[\wtilde{Y}]$ and $|Y - \wtilde{Y} - \mu| \le_{0} \frac{h}{2}$, for some $\mu$ with $|\mu| \le \frac{h}{2}$.
\end{proposition}

\section{Two-Stage Composition Algorithm}\label{sec:two-stage-compose}

Our two-stage algorithm for the case of $k$-fold self-composition is given in \cref{alg:two-stage-self-compose}.  We assume $k = k_1 \cdot k_2 + r$ where $k_1, k_2 \in \mathbb{Z}_{> 0}$, $r < k_1$, and $k_1 \approx k_2 \approx \sqrt{k}$, which for instance can be achieved by taking $k_1 = \lfloor\sqrt{k}\rfloor$, $k_2 = \floor{k/k_1}$, and $r = k - k_1 \cdot k_2$. 

The algorithm implements the {\em circular convolution} $\oplus_L$ using Fast Fourier Transform (FFT). For any $L$ and $x \in \bbR$, we define $x \mod{2L} = x - 2Ln$ where $n \in \bbZ$ such that $x - 2Ln \in (-L, L]$. Given $x, y \in \bbR$ the \emph{circular addition} is defined as 
\[
x \oplus_L y ~:=~ x+y \mod{2L}.
\]
Similarly, for random variables $X, Y$, we define $X \oplus_L Y$ to be their convolution modulo $2L$ and $Y^{\oplus_L k}$ to be the $k$-fold convolution of $Y$ modulo $2L$.

Observe that $\DiscretizeRV$ with mesh size $h$ and truncation parameter $L$ runs in time $O(L/h)$, assuming an $O(1)$-time access to $\CDF_Y(\cdot)$. The FFT convolution step runs in time $O\inparen{\frac{L_i}{h_i} \log \frac{L_i}{h_i}}$; thus the overall running time of $\TwoStageSelfComposePRV$ is
\begin{align*}
	\textstyle O\inparen{\frac{L_1}{h_1} \log\inparen{\frac{L_1}{h_1}} + \frac{L_2}{h_2} \log\inparen{\frac{L_2}{h_2}}}\,.
\end{align*}

\begin{algorithm}[t]
	\caption{$\TwoStageSelfComposePRV$}
	\label{alg:two-stage-self-compose}
	\begin{algorithmic}
		\STATE {\bfseries Input:} PRV $Y$, number of compositions $k = k_1 \cdot k_2 + r$ (for $r < k_1$), mesh sizes $h_1 \le h_2$, truncation parameters $L_1 \le L_2$, where each $L_i \in h_i \cdot (\frac{1}{2} + \bbZ_{>0})$.
		\STATE {\bfseries Output:} PDF of an approximation $Y_{2}$ for $Y^{\oplus k}$. $Y_{2}$ will be supported on $\mu + (h_2\bbZ \cap [-L_2, L_2])$ for some $\mu \in \insquare{0, \frac{h_2}{2}}$.
		\STATE
		\STATE $Y_0 \gets Y |_{|Y| \le L_1}$ \hfill \texttt{ \small $\triangleright$ $Y$ conditioned on $|Y| \le L_1$}
		\STATE $\wtilde{Y}_{0} \gets \DiscretizeRV(Y_{0}, h_{1}, L_{1})$
		\STATE $Y_{1} \gets \wtilde{Y}_{0}^{\oplus_{L_{1}} k_1}$ 
		\hfill \texttt{\small $\triangleright$ $k_1$-fold FFT convolution}\\
		\STATE $\wtilde{Y}_{1} \gets \DiscretizeRV(Y_{1}, h_{2}, L_{2})$ 
		\STATE $Y_{2} \gets \wtilde{Y}_{1}^{\oplus_{L_2} k_2}$ \hfill \texttt{\small $\triangleright$ $k_2$-fold FFT convolution}
		\STATE $Y_{r} \gets \wtilde{Y}_{0}^{\oplus_{L_{1}} r}$ \hfill \texttt{\small $\triangleright$ $r$-fold FFT convolution}
		\STATE $\wtilde{Y}_{r} \gets \DiscretizeRV(Y_{r}, h_{2}, L_{2})$
		\RETURN $Y_2 \oplus_{L_2} \wtilde{Y}_r$  \hfill \texttt{\small $\triangleright$ FFT convolution}
	\end{algorithmic}
\end{algorithm}

The approximation guarantees provided by our two-stage algorithm are captured in the following theorem.  For convenience, we assume that $k$ is a perfect square (we set $k_1 = k_2 = \sqrt{k}$).  The complete proof is in Section~\ref{sec:2stageanalysis}.
\begin{theorem}\label{thm:two-stage-self-compose-error}
For any PRV $Y$, the approximation $Y_{2}$ returned by $\TwoStageSelfComposePRV$ satisfies
\[
	|Y^{\oplus k} - Y_{2}| \le_{\deltaerror} \epserror,
\]
when invoked with $k_1 = k_2 = k^{0.5}$ (assumed to be an integer) and parameters given below (using $\eta := \frac{\deltaerror}{(8 \sqrt{k} + 16)}$)
\[
\textstyle h_1 := \frac{\epserror}{k^{0.5} \sqrt{2 \log \frac{1}{\eta}}}\,, \quad h_2 := \frac{\epserror}{k^{0.25}\sqrt{2 \log \frac{1}{\eta}}}
\]
\begin{align*}
L_1 &\textstyle~\ge~ \max\set{\begin{matrix}
		\eps_{Y}\inparen{\frac{\epserror\deltaerror}{16 k^{1.25}}} %
		, \\
		\eps_{Y^{\oplus \sqrt{k}}}\inparen{\frac{\epserror\deltaerror}{64 k^{0.75}}} %
	\end{matrix}}  + \frac{\epserror}{k^{0.25}} \\
L_2 &\textstyle~\ge~ \max\set{\eps_{Y^{\oplus k}}\inparen{\frac{\epserror\deltaerror}{16}} + 2\epserror, L_1}.
\end{align*}
\end{theorem}

In terms of a concrete running time bound, \cref{thm:two-stage-self-compose-error} implies:%
\begin{corollary}\label{cor:informal-two-stage-self-compose-error}
For any DP algorithm $\calM$, the privacy curve $\delta_{\calM^{\circ k}}(\eps)$ of the $k$-fold (adaptive) composition of $\calM$ can be approximated in time
$
\wtilde{O}\inparen{\frac{\epsupper \cdot k^{0.25} \cdot \sqrt{\log(k/\deltaerror)}}{\epserror}},
$
where $\epserror$ is the additive error in $\eps$, $\deltaerror$ is the additive error in $\delta$, and $\epsupper$ is an upper bound on
\begin{align*}
\max\set{\inmat{
\eps_{Y^{\oplus k}}(\frac{\epserror\deltaerror}{16}),\\[1mm]
\sqrt[4]{k}\cdot \eps_{Y^{\oplus \sqrt{k}}}\inparen{\frac{\epserror\deltaerror}{64k^{0.75}}},\\[1mm]
\sqrt[4]{k}\cdot \eps_Y\inparen{\frac{\epserror\deltaerror}{16k^{1.25}}}
}} + \epserror.
\end{align*}
\end{corollary}

In many practical regimes of interest, $\epsupper/\epserror$ is a constant. For ease of exposition in the following, we assume that $\epserror$ is a small constant, e.g. $0.1$ and suppress the dependence on $\epserror$.
Suppose the original mechanism $\calM$ underlying $Y$ satisfies $(\eps = \frac{1}{\sqrt{k \cdot \log(1/\deltaerror)}}, \delta = o_k\inparen{\frac{1}{k^{1.25}}})$-DP.  Then by  advanced composition~\cite{dwork2010boosting}, we have that $\calM^{\circ t}$ satisfies $(\eps \sqrt{2t \log \frac{1}{\delta'}} + 2t\eps (e^{\eps} - 1), t\delta + \delta')$-DP. If $t\delta + \delta' \lesssim \frac{t\deltaerror}{k^{1.25}}$, then we have that $\eps_{Y^{\oplus t}}\inparen{\frac{t\deltaerror}{k^{1.25}}} \lesssim \sqrt{\frac{t}{k} \ln \frac{k}{t\deltaerror}}$. Instantiating this with $t = 1, \sqrt{k}$, and $k$ gives us that $\epsupper$ is at most a constant.

Note that, to set the value of $L_1$ and $L_2$, we do not need the exact value of $\eps_{Y^{\oplus k}}$ (or $\eps_{Y^{\oplus \sqrt{k}}}$ or $\eps_Y$). We only need an upper bound on $\eps_{Y^{\oplus k}}$, which can often be obtained by using the RDP accountant or some other method.

For the case when $k$ is not a perfect square, using a similar analysis, it is easy to see that the approximation error would be no worse than the error in $k_2 (k_1 + 1)$ self-compositions. The running time increases by a constant because of the additional step of $r$-fold convolution to get $Y_r$ and the final convolution step to get $Y_2 \oplus_{L_2} \wtilde{Y}_r$; however this does not affect the asymptotic time complexity of the algorithm. Moreover, as seen in \cref{sec:experiments}, even with this additional cost, $\TwoStageSelfComposePRV$ is still faster than the algorithm in \citet{gopi2021numerical}.

\subsection{Analysis}
\label{sec:2stageanalysis}

In this section we establish Theorem~\ref{thm:two-stage-self-compose-error}.  The proof proceeds are follows.  We establish coupling approximations between consecutive random variables in the following sequence:
\[ Y^{\oplus k_1 k_2},\ \ Y_{0}^{\oplus k_1 k_2},\ \ \wtilde{Y}_{0}^{\oplus k_1 k_2},\ \ Y_1^{\oplus k_2},\ \ \wtilde{Y}_1^{\oplus k_2}, Y_2\,, \]
and then apply \Cref{lem:coupling-properties}\eqref{item:coupling-triangle}.

To establish each coupling approximation, we use Lemmas~\ref{lem:coupling-properties}\eqref{item:coupling-composition} and~\ref{lem:coupling-properties}\eqref{item:coupling-concentration}.

\paragraph{\boldmath Coupling $\inparen{Y^{\oplus k_1 k_2}, \ Y_0^{\oplus k_1 k_2}}$.}
Since $\dTV(Y, Y_0) = \Pr[|Y| > L_1] =: \delta_0$, we have from \cref{lem:coupling-properties}\eqref{item:coupling-composition} that
\begin{align}
	|Y^{\oplus k_1k_2} - Y_0^{\oplus k_1k_2}| &~\le_{k_1k_2 \delta_0}~ 0\,.\label{eq:2stg_base-case}
\end{align}

\paragraph{\boldmath Coupling $\inparen{Y_0^{\oplus k_1k_2}, \ \wtilde{Y}_{0}^{\oplus k_1k_2}}$ and $\inparen{Y_1^{\oplus k_2}, \ \wtilde{Y}_{1}^{\oplus k_2}}$.}
We have from \cref{claim:discretize-prv-prop} that $\Ex[Y_0] = \Ex[\wtilde{Y}_0]$ and that $|Y_0 - \wtilde{Y}_0 - \mu| \le_0 \frac{h_{1}}{2}$. Thus, by applying \cref{lem:coupling-properties}\eqref{item:coupling-concentration}, we have that (for any $\eta$; to be chosen later)
\begin{align}
\inabs{Y_0^{\oplus k_1} - \wtilde{Y}_{0}^{\oplus k_1}} &\textstyle~\le_{\eta}~ h_{1} \sqrt{\frac{k_1}{2} \log \frac{2}{\eta}}, \label{eq:2stg_re-discretize-1a}\\
\inabs{Y_0^{\oplus k_1k_2} - \wtilde{Y}_{0}^{\oplus k_1k_2}} &\textstyle~\le_{\eta}~ h_{1} \sqrt{\frac{k_1k_2}{2} \log \frac{2}{\eta}}. \label{eq:2stg_re-discretize-1b}
\end{align}
Similarly, we have that
\begin{align}
	\inabs{Y_1^{\oplus k_2} - \wtilde{Y}_{1}^{\oplus k_2}} &\textstyle~\le_{\eta}~ h_{2} \sqrt{\frac{k_2}{2} \log \frac{2}{\eta}}. \label{eq:2stg_re-discretize-2}
\end{align}

\paragraph{\boldmath Coupling $\inparen{\wtilde{Y}_0^{\oplus k_1k_2}, \ Y_{1}^{\oplus k_2}}$ and $\inparen{\wtilde{Y}_1^{\oplus k_2}, \ Y_{2}}$.}
Since $\dTV(\wtilde{Y}_0^{\oplus k_1}, \wtilde{Y}_0^{\oplus_{L_{1}} k_1}) \le \Pr\insquare{\inabs{\wtilde{Y}_0^{\oplus k_1}} > L_{1}} =: \delta_{1}$, and $Y_1 = Y_0^{\oplus_{L_1} k_1}$, it holds via \cref{lem:coupling-properties}\eqref{item:coupling-composition} that
\begin{align}
	\inabs{\wtilde{Y}_0^{\oplus k_1k_2} - Y_{1}^{\oplus k_2}} \le_{k_2 \delta_{1}} 0\,.\label{eq:2stg_wrap-around-1}
\end{align}
Similarly, for $\delta_2 := \Pr\insquare{\inabs{\wtilde{Y}_1^{\oplus k_2}} > L_{2}}$, we have that
\begin{align}
	\inabs{\wtilde{Y}_1^{\oplus k_2} - Y_{2}} \le_{\delta_{2}} 0\,.\label{eq:2stg_wrap-around-2}
\end{align}

\paragraph{Towards combining the bounds.}
Combining \cref{eq:2stg_base-case,eq:2stg_re-discretize-1b,eq:2stg_wrap-around-1,eq:2stg_re-discretize-2,eq:2stg_wrap-around-2} using \cref{lem:coupling-properties}\eqref{item:coupling-triangle}, we have that
\begin{align}
	& \inabs{Y^{\oplus k_1k_2} - Y_2} ~\le_{\deltaerror}~ \epserror, \nonumber\\
	\text{where } \ \ \deltaerror &\textstyle~=~ 2\eta + k_1k_2 \delta_0 + k_2 \delta_1 + \delta_2\,, \label{eq:2stg_delta-star}\\
	\text{and } \ \ \epserror &\textstyle~=~ \inparen{h_1 \sqrt{k_1 k_2} + h_2 \sqrt{k_2} } \sqrt{\frac12 \log \frac{2}{\eta}} \label{eq:2stg_eps-star}\,.
\end{align}
We set $h_1 := \frac{\epserror}{\sqrt{2k_1k_2 \log \frac{2}{\eta}}}$ and $h_2 := \frac{\epserror}{\sqrt{2k_2 \log \frac{2}{\eta}}}$. The key step remaining is to bound $\delta_0$, $\delta_1$, and $\delta_2$ in terms of $h_i$'s, $L_i$'s, and $\eta_i$'s.

To do so, we use \Cref{lem:tail-bounds} for $\alpha_i$'s to be chosen later.

\paragraph{\boldmath Bounding $\delta_0$.} We have $\delta_0 := \Pr\insquare{\inabs{Y} > L_1}$ and hence
\begin{align*}
\delta_0 &\textstyle~\le~ \frac{4}{\alpha_0} \delta_Y(L_1 - \alpha_0). 
\end{align*}

\paragraph{\boldmath Bounding $\delta_1$.} For $\wtilde{h}_1 := h_1 \sqrt{\frac{k_1}{2} \log \frac{2}{\eta}} = \frac{\epserror}{2\sqrt{k_2}}$, we have
\begin{align*}
\delta_1
&\textstyle~:=~ \Pr\insquare{\inabs{\wtilde{Y}_0^{\oplus k_1}} > L_1} \nonumber\\
&\textstyle~\le~ \Pr\insquare{\inabs{\wtilde{Y}_0^{\oplus k_1} - Y_0^{\oplus k_1}} > \wtilde{h}_1} \nonumber\\
&\textstyle\quad +~ \Pr\insquare{\inabs{Y_0^{\oplus k_1}} > L_1 - \wtilde{h}_1}\nonumber\\
&\textstyle~\le~ \eta + \Pr\insquare{\inabs{Y^{\oplus k_1}} > L_1 - \wtilde{h}_1}\nonumber\\
\Longrightarrow \quad \delta_1 &\textstyle~\le~ \eta + \frac{4}{\alpha_1} \delta_{Y^{\oplus k_1}}\inparen{L_1 - \frac{\epserror}{2\sqrt{k_2}} - \alpha_1}, 
\end{align*}
where the second inequality uses \cref{eq:2stg_re-discretize-1a} and the third inequality uses the fact that the tails of $Y^{\oplus k_1}$ are only heavier than the tails of $Y_0^{\oplus k_1}$ since $Y_0$ is a truncation of $Y$.

\paragraph{\boldmath Bounding $\delta_2$.} First, we combine \cref{eq:2stg_re-discretize-1b,eq:2stg_re-discretize-2,eq:2stg_wrap-around-1} to get (recall $\epserror$ in \cref{eq:2stg_eps-star})
\begin{align*}
	\inabs{Y_0^{\oplus k_1k_2} - \wtilde{Y}_1^{\oplus k_2}}
	&\textstyle~\le_{2\eta + k_2 \delta_1}~ \epserror. %
\end{align*}
Using this, we get
\begin{align*}
\delta_2
&\textstyle~:=~ \Pr\insquare{\inabs{\wtilde{Y}_1^{\oplus k_2}} > L_2} \nonumber\\
&\textstyle~\le~ \Pr\insquare{\inabs{\wtilde{Y}_1^{\oplus k_2} - Y_0^{\oplus k_1k_2}} > \epserror}\nonumber\\
&\textstyle\qquad +~ \Pr\insquare{\inabs{Y_0^{\oplus k_1k_2}} > L_2 - \epserror}\nonumber\\
&\textstyle~\le~ 2\eta + k_2 \delta_1 + \Pr\insquare{\inabs{Y^{\oplus k_1k_2}} > L_2 - \epserror}\nonumber\\
&\textstyle~\le~ 2\eta + k_2 \delta_1 + \frac{4}{\alpha_2}\delta_{Y^{\oplus k_1k_2}}\inparen{L_2 - \epserror - \alpha_2}\nonumber\\
\Longrightarrow \quad \delta_2 &\textstyle~\le~ (k_2 + 2)\eta + k_2 \cdot \frac{4}{\alpha_1}\delta_{Y^{\oplus k_1}}(L_1 - \frac{\epserror}{2\sqrt{k_2}} - \alpha_1) \nonumber\\
&\textstyle\qquad +~ \frac{4}{\alpha_2} \delta_{Y^{\oplus k_1k_2}}\inparen{L_2 - \epserror - \alpha_2} ,
\end{align*}
where we use in the third step that tails of $Y^{\oplus k_1k_2}$ are only heavier than tails of $Y_0^{\oplus k_1k_2}$ since $Y_0$ is a truncation of $Y$.

\paragraph{Putting it all together.}
Plugging in these bounds for $\delta_0$, $\delta_1$, and $\delta_2$ in \cref{eq:2stg_delta-star}, we get that
\begin{align}
& \inabs{Y^{\oplus k_1k_2} - Y_2} ~\le_{\deltaerror}~ \epserror, \nonumber\\[1mm]
\text{where } \ \ \deltaerror &\textstyle~\le~ (2k_2 + 4) \eta + k_1k_2 \cdot \frac{4}{\alpha_0}\delta_Y(L_1 - \alpha_0) \nonumber\\
&\textstyle\qquad +~ 2 k_2 \cdot \frac{4}{\alpha_1}\delta_{Y^{\oplus k_1}}(L_1 - \frac{\epserror}{2\sqrt{k_2}} - \alpha_1)\nonumber\\
&\textstyle\qquad +~ \frac{4}{\alpha_2}\delta_{Y^{\oplus k_1k_2}}\inparen{L_2 - \epserror - \alpha_2}\,, \label{eq:2stg_delta-star-final}\\
\text{and } \ \ \epserror &\textstyle~=~ \inparen{h_1 \sqrt{k_1 k_2} + h_2 \sqrt{k_2} } \sqrt{2 \log \frac{2}{\eta}} \nonumber.
\end{align}

Thus, we can set parameters $L_1$, $L_2$, and $\eta$ as follows such that each of the four terms in \cref{eq:2stg_delta-star-final} is at most $\deltaerror/4$ to satisfy the above:
\begin{align*}
\eta &\textstyle~=~ \frac{\deltaerror}{8k_2 + 16}, \\
L_1 &\textstyle~\ge~ \max\set{\begin{matrix}
		\eps_{Y}\inparen{\frac{\alpha_0\deltaerror}{16k_1k_2}} + \alpha_0, \\
		\eps_{Y^{\oplus k_1}}\inparen{\frac{\alpha_1\deltaerror}{32k_2}} + \frac{\epserror}{2\sqrt{k_2}} + \alpha_1
	\end{matrix}}, \\
L_2 &\textstyle~\ge~ \max\set{\eps_{Y^{\oplus k_1k_2}}\inparen{\frac{\alpha_2\deltaerror}{16}} + \epserror + \alpha_2, L_1}.
\end{align*}

Setting $k_1 = k_2 = \sqrt{k}$ (assumed to be integers) and $\alpha_0 = \frac{\epserror}{\sqrt{k_2}}$, $\alpha_1 = \frac{\epserror}{2\sqrt{k_2}}$ and $\alpha_2 = \epserror$, completes the proof of \cref{thm:two-stage-self-compose-error}.

\paragraph{Runtime analysis.} As argued earlier, the total running time of $\TwoStageSelfComposePRV$ is given as $\textstyle O\inparen{\frac{L_1}{h_1}\log \frac{L_1}{h_1} + \frac{L_2}{h_2} \log \frac{L_2}{h_2}}$. Substituting in the bounds for $L_1$, $L_2$, $h_1$, and $h_2$, we get a final running time of 
\[
\wtilde{O}\inparen{\frac{\epsupper \cdot k^{0.25} \cdot \sqrt{\log(k/\deltaerror)}}{\epserror}}.
\]
where $\epsupper$ is as in \Cref{cor:informal-two-stage-self-compose-error}.

\subsection{Heterogeneous Compositions}
\cref{alg:two-stage-self-compose} can be easily generalized to handle {\em heterogeneous composition} of $k$ different mechanisms, with a running time blow up of $k$ over the homogeneous case. We defer the details to \cref{apx:two-stage-extensions}.

\section{\boldmath Experimental Evaluation of $\TwoStageSelfComposePRV$}\label{sec:experiments}

\def\figheight{4cm}
\begin{figure*}[t]
	\centering
	\begin{subfigure}[t]{0.27\textwidth}
		\centering
		\includegraphics[height=4cm]{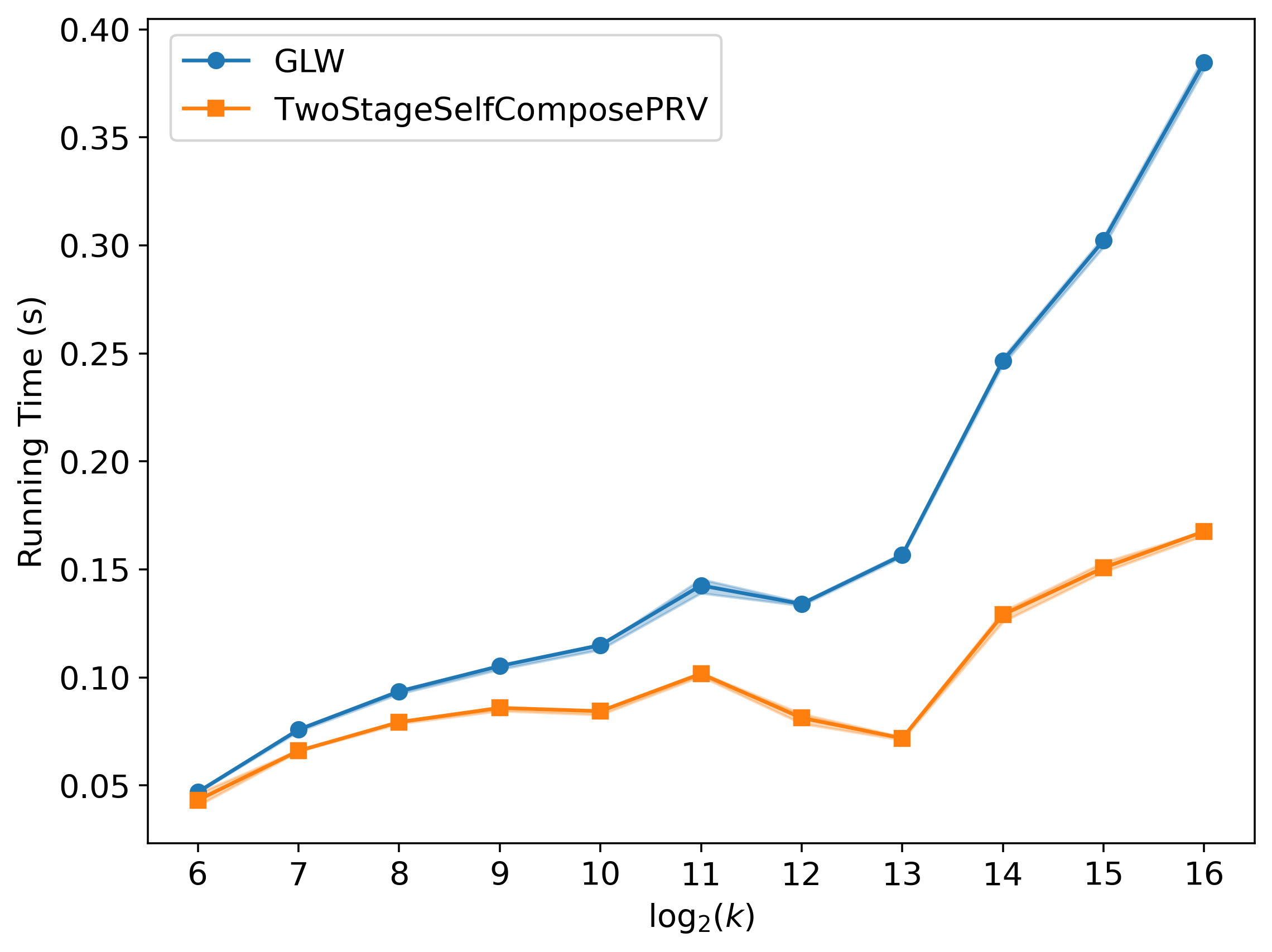}
		\caption{Running times (average over 20 runs); shaded region indicates 20th--80th percentiles}
		\label{fig:laplace_runtimes}
	\end{subfigure}
	\hfill 
	\begin{subfigure}[t]{0.33\textwidth}
		\centering
		\includegraphics[height=4cm]{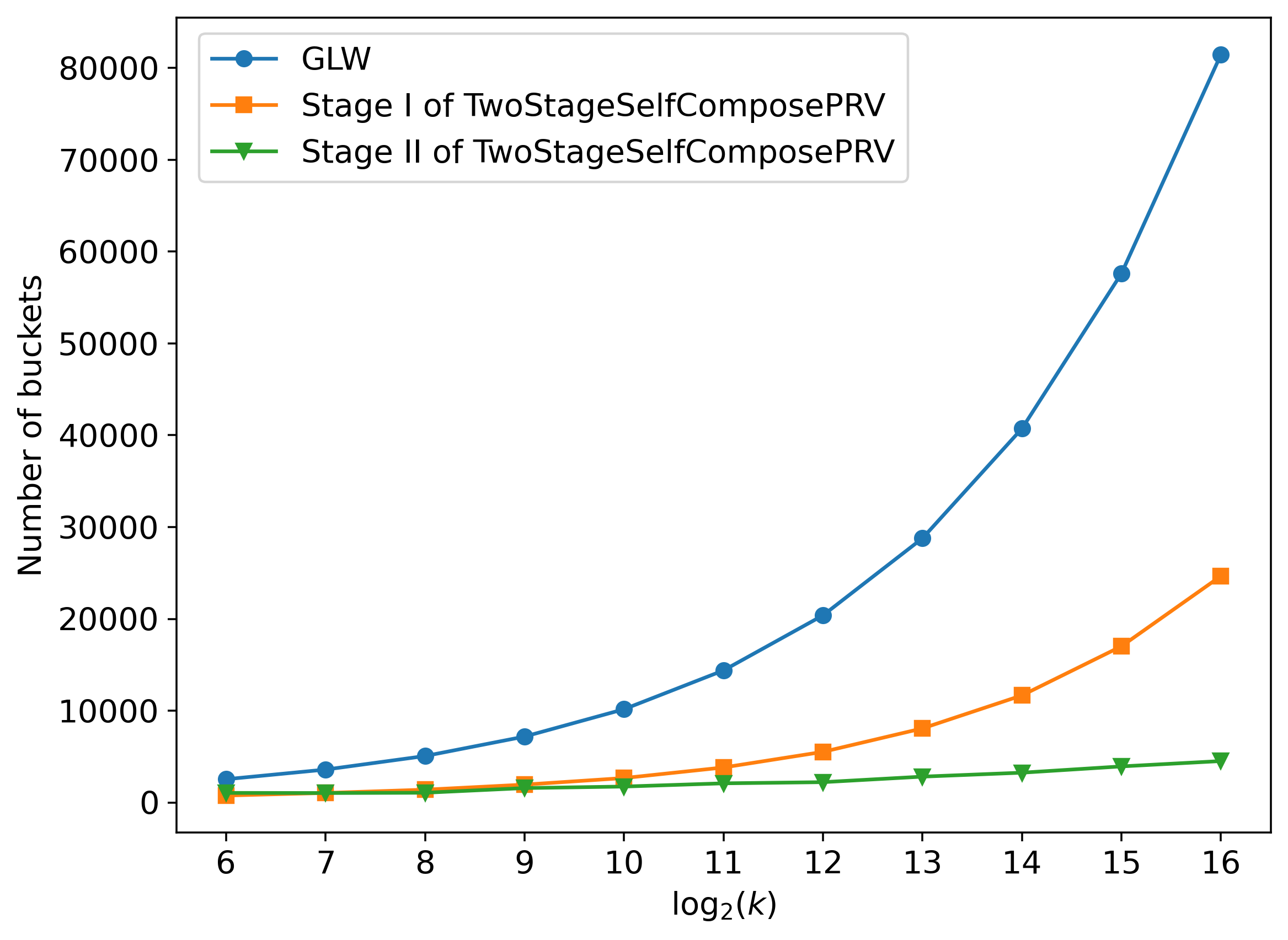}
		\caption{Number of discretization buckets}
		\label{fig:laplace_buckets}
	\end{subfigure}
	\begin{subfigure}[t]{0.33\textwidth}
		\centering
		\includegraphics[height=4.15cm]{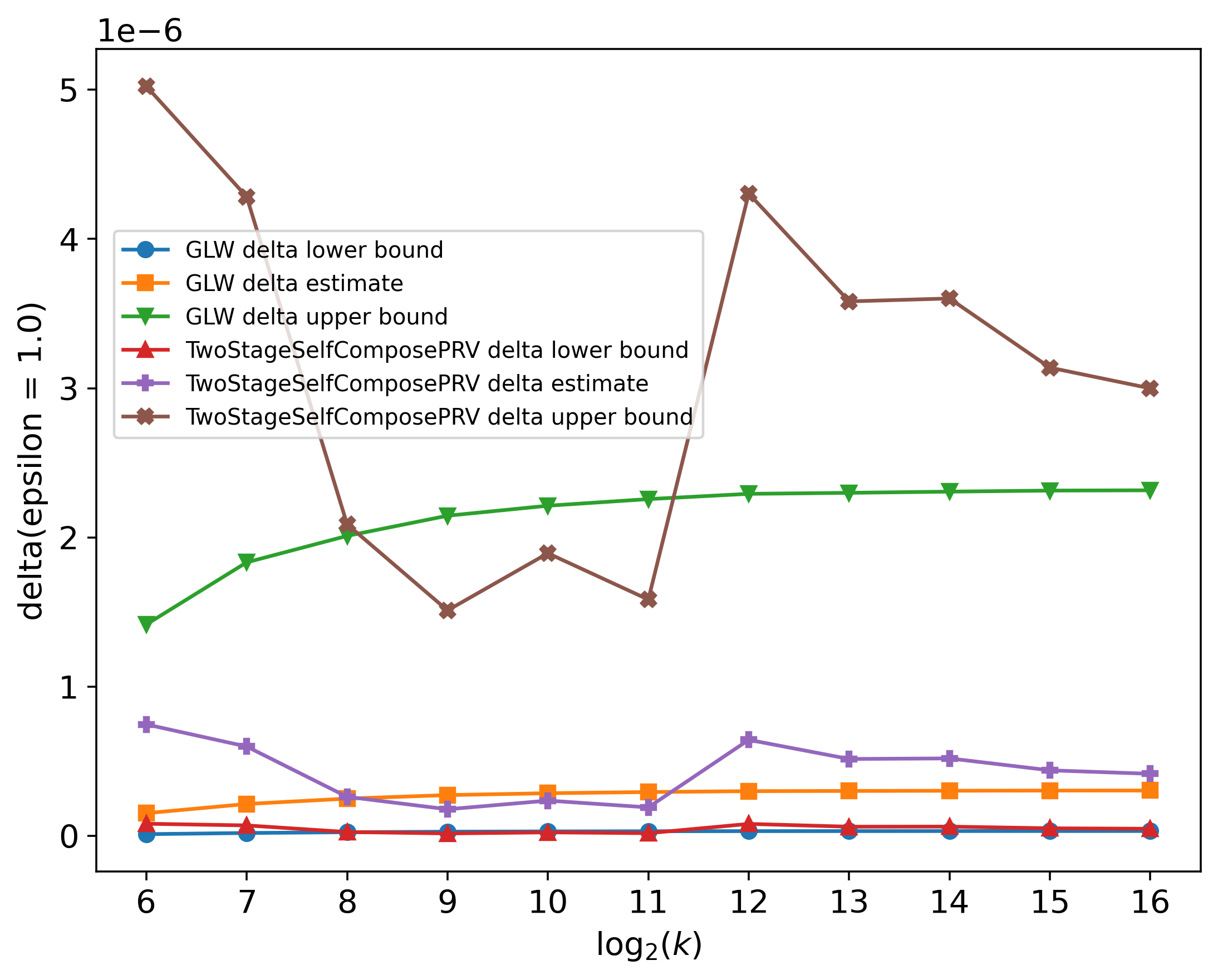}
		\caption{Delta estimates}
		\label{fig:laplace_delta}
	\end{subfigure}
	\caption{Compositions of the Laplace mechanism.}
	\label{fig:laplace}
\end{figure*}

\begin{figure*}[t]
	\centering
     \begin{subfigure}[t]{0.27\textwidth}
		\centering
		\includegraphics[height=4cm]{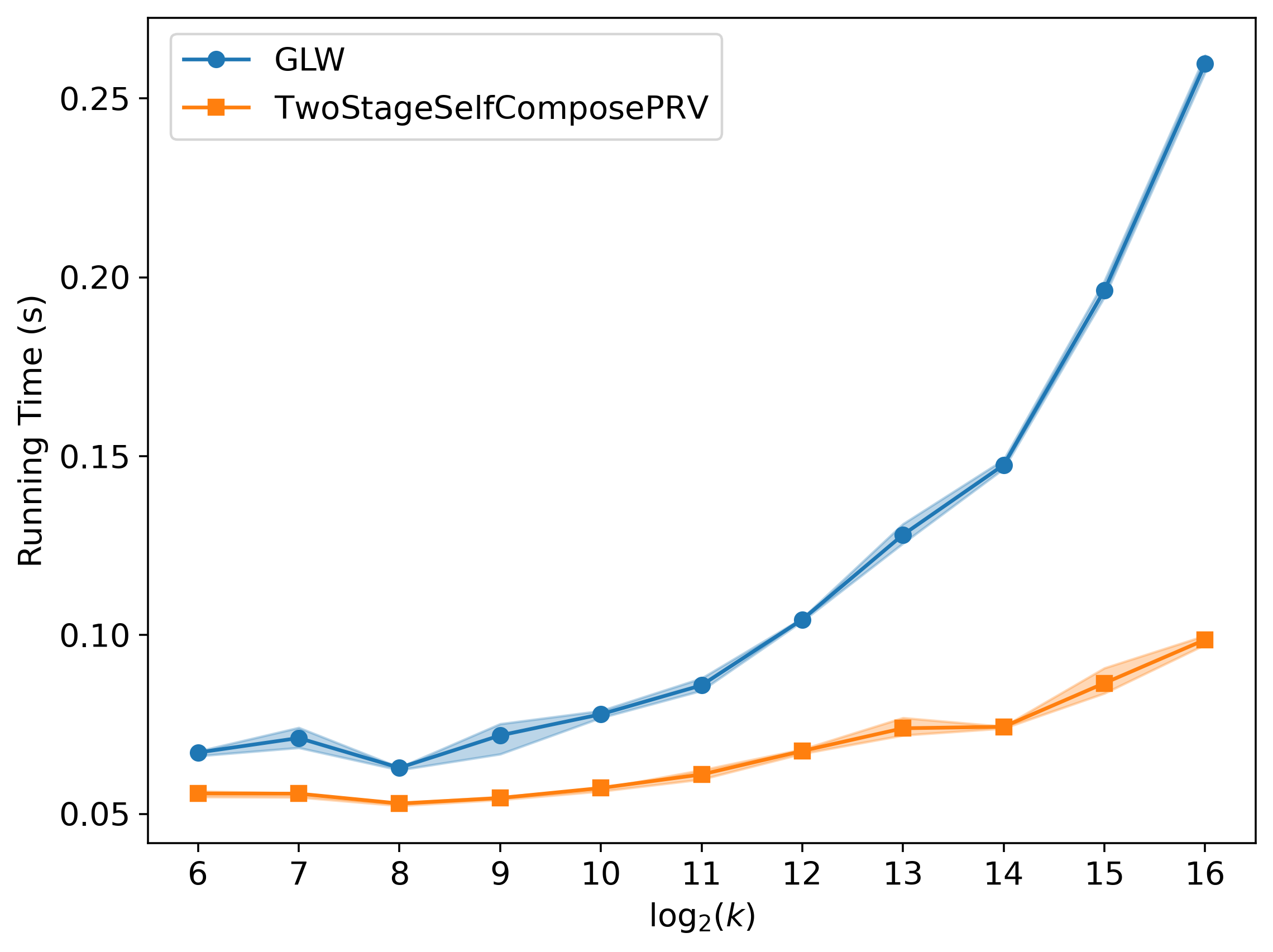}
		\caption{Running times (average over 20 runs); shaded region indicates 20th--80th percentiles}
		\label{fig:gaussian_runtimes}
	\end{subfigure}
	\hfill 
	\begin{subfigure}[t]{0.33\textwidth}
		\centering
		\includegraphics[height=4cm]{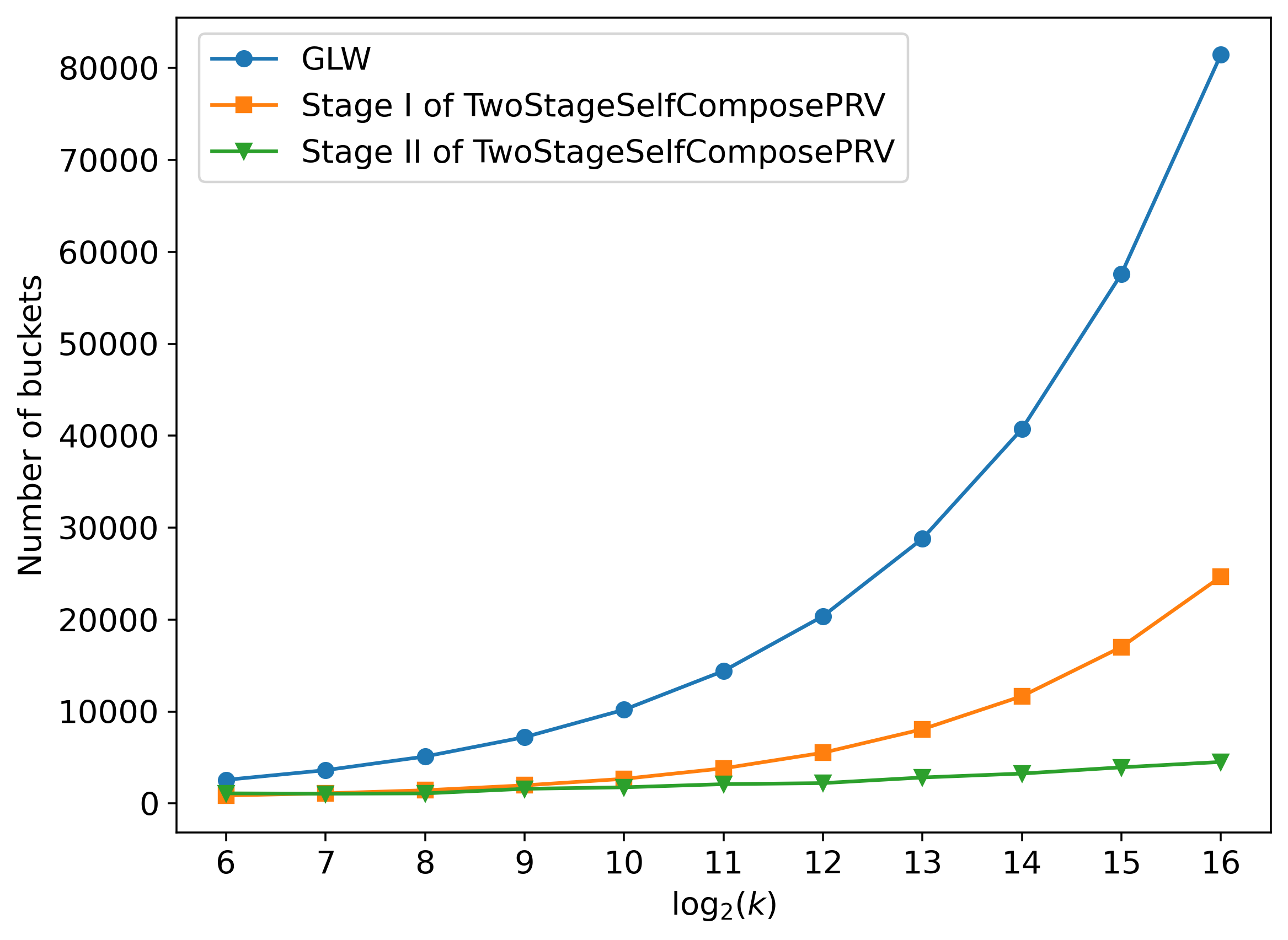}
		\caption{Number of discretization buckets}
		\label{fig:gaussian_buckets}
	\end{subfigure}
	\begin{subfigure}[t]{0.33\textwidth}
		\centering
		\includegraphics[height=4.15cm]{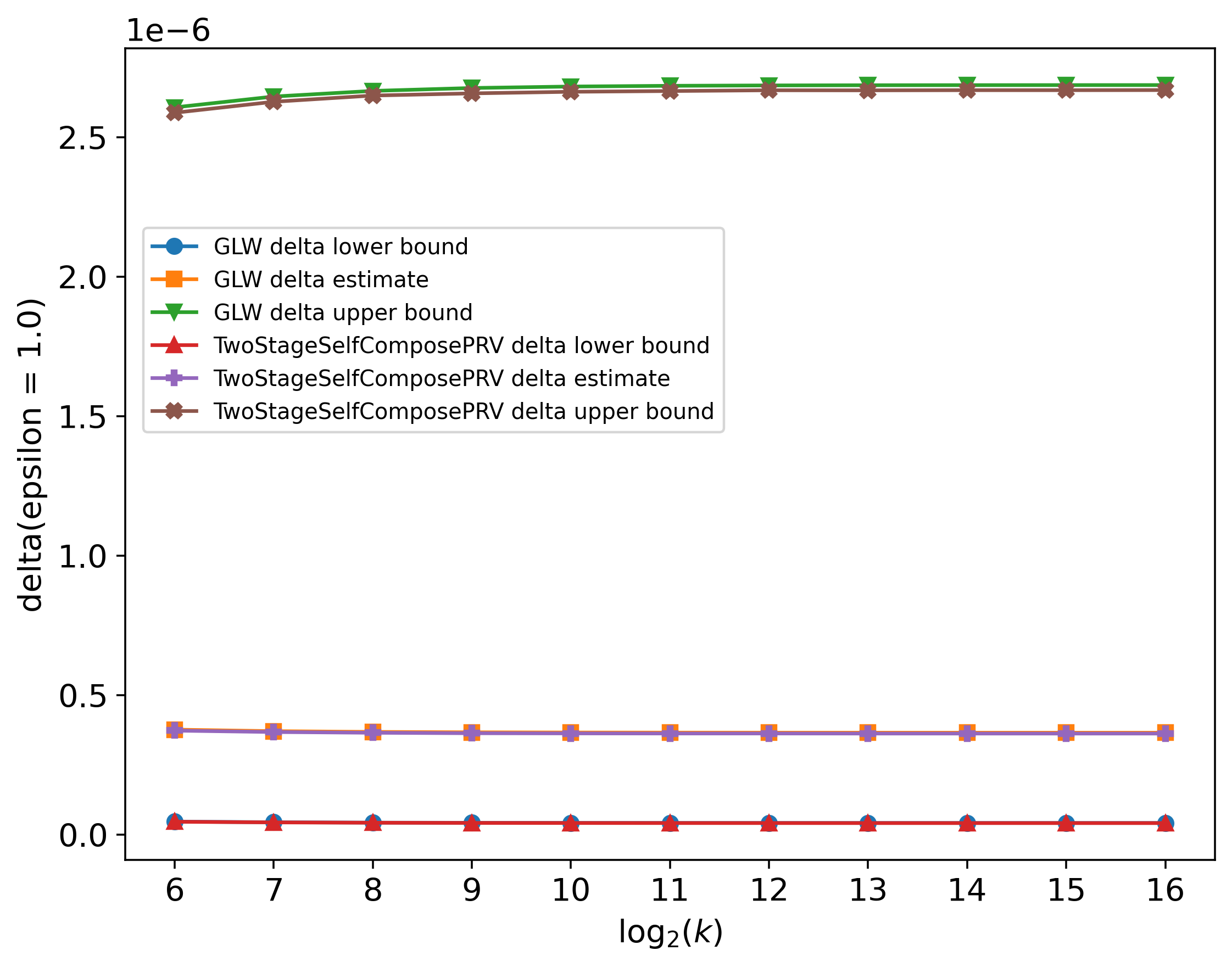}
		\caption{Delta estimates}
		\label{fig:gaussian_delta}
	\end{subfigure}
	\caption{Compositions of Poisson subsampled (with probability $\gamma=0.2$) Gaussian mechanism.}
	\label{fig:subsamp-gaussian}
\end{figure*}

\begin{figure*}
    \centering
	\begin{subfigure}[t]{0.45\textwidth}
		\centering
		\includegraphics[height=4.5cm]{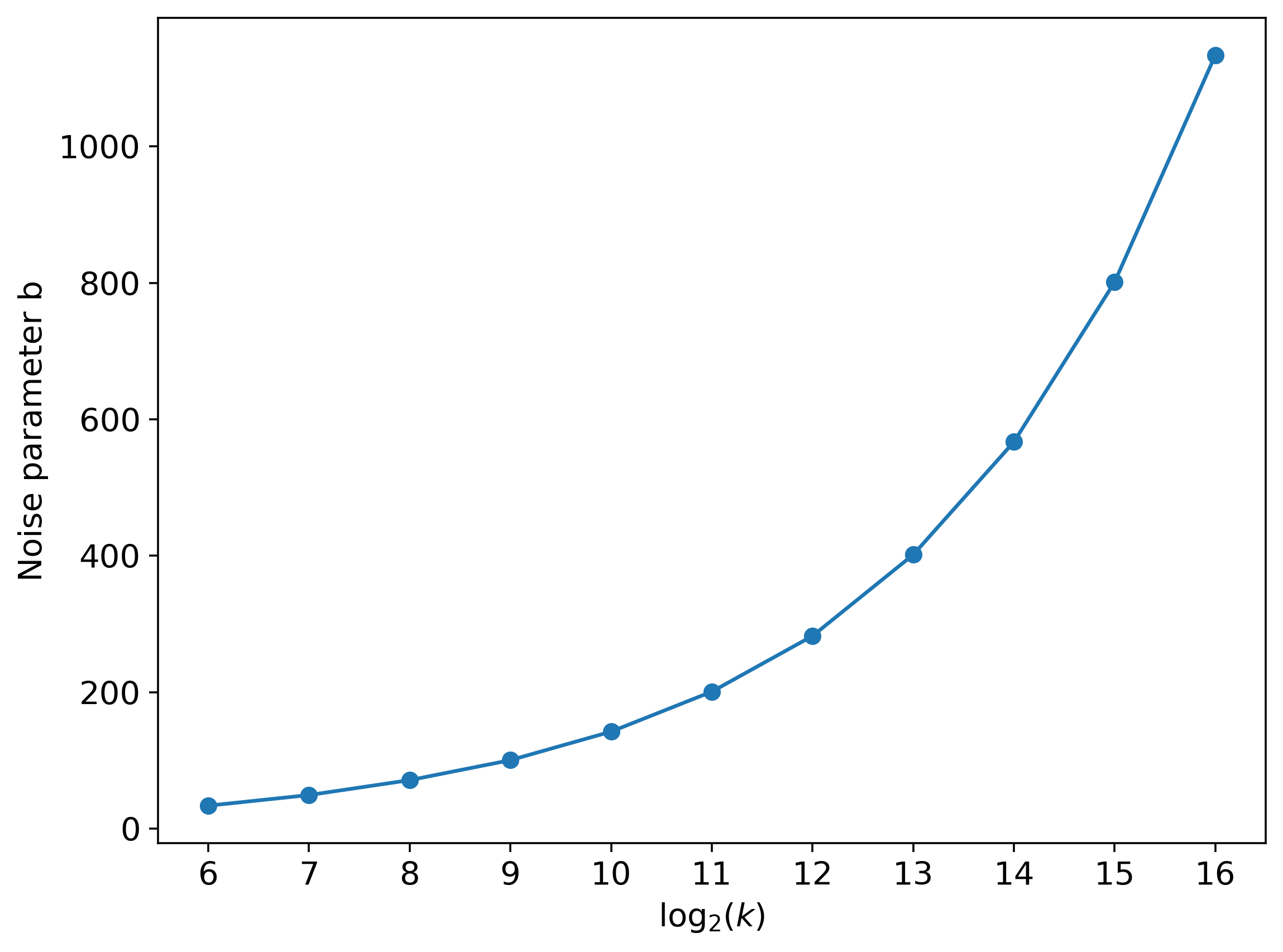}
		\caption{For Laplace mechanism (\Cref{fig:laplace}).}
		\label{fig:laplace_noise}
	\end{subfigure}
	\hfill 
	\begin{subfigure}[t]{0.45\textwidth}
		\centering
		\includegraphics[height=4.5cm]{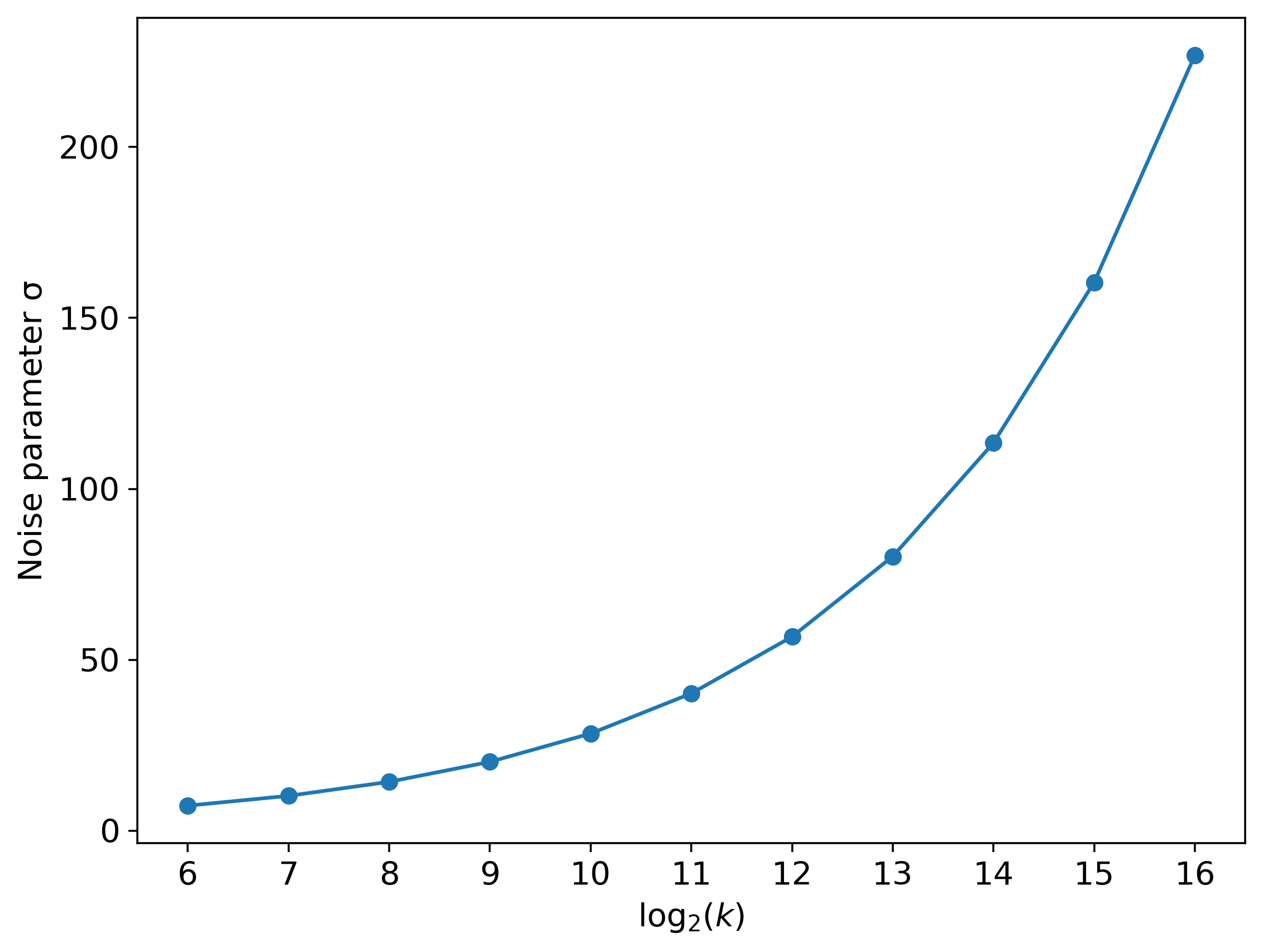}
		\caption{For Poisson subsampled Gaussian mechanism (\Cref{fig:subsamp-gaussian}).}
		\label{fig:gaussian_noise}
	\end{subfigure}
	\caption{Noise parameters used for experiments.}
	\label{fig:noise_params}
\end{figure*}

We empirically evaluate $\TwoStageSelfComposePRV$ on the tasks of self-composing two kinds of mechanism acting on dataset of real values $x_1, \ldots, x_n$ as
\begin{itemize}
\item Laplace Mechanism : returns $\sum_i x_i$ plus a noise drawn from $L(0, b)$ given by the $\PDF$ $P(x) = e^{-|x|/b}/2b$.
\item Poisson Subsampled Gaussian Mechanism with probability $\gamma$: Subsamples a random subset $S$ of indices by including each index independently with probability $\gamma$. Returns $\sum_{i \in S} x_i$ plus a noise drawn from the Gaussian distribution $\calN(0, \sigma^2)$.
\end{itemize}
Both these mechanisms are highly used in practice. For each mechanism, we compare against the implementation by \citet{gopi2021numerical}\footnote{\scriptsize\texttt{github.com/microsoft/prv\_accountant}} (referred to as \textsf{GLW}) on three fronts: (i) the running time of the algorithm, (ii) the number of discretized buckets used, and (iii) the final estimates on $\delta_{Y^{\oplus k}}(\eps)$ which includes comparing lower bound $\delta_{Y_2}(\eps + \epserror) - \deltaerror$, estimates $\delta_{Y_2}(\eps)$ and upper bounds $\delta_{Y_2}(\eps - \epserror) + \deltaerror$. We use $\epserror = 0.1$ and $\deltaerror = 10^{-10}$ in all the experiments.

We run each algorithm for a varying number $k$ of self-compositions of a basic mechanism. The noise parameter of basic mechanism is chosen such that the final $\delta(\eps)$ value after $k$-fold composition is equal to $10^{-6}$ for each value of $k$.\footnote{these values were computed using the Google DP accountant {\scriptsize\texttt{github.com/google/differential-privacy/tree/main/python}}.} The exact choice of noise parameters used are shown in \Cref{fig:noise_params}.

The comparison for the Laplace mechanism is shown in \cref{fig:laplace} and for the subsampled Gaussian mechanism is shown in \cref{fig:subsamp-gaussian}. In terms of accuracy we find that for the same choice of $\epserror$ and $\deltaerror$, the estimates returned by $\TwoStageSelfComposePRV$ are nearly identical to the estimates returned by \textsf{GLW} for the subsampled Gaussian mechanism. On the other hand, the estimates for Laplace mechanism returned by both algorithms are similar and consistent with each other, but strictly speaking, incomparable with each other.%

\section{Multi-Stage Recursive Composition}\label{sec:recursive-compose}

We extend the approach in $\TwoStageSelfComposePRV$ to give a multi-stage algorithm (\cref{alg:self-compose-prv}), presented only when $k$ is a power of $2$ for ease of notation. Similar to the running time analysis of $\TwoStageSelfComposePRV$, the running time of $\RecursiveSelfComposePRV$ is given as
\begin{align*}
	O\inparen{\sum_{i=1}^t \frac{L_i}{h_i} \log\inparen{\frac{L_i}{h_i}}},
\end{align*}
assuming an $O(1)$-time access to $\CDF_Y(\cdot)$.

\begin{algorithm}[t]
\caption{$\RecursiveSelfComposePRV$}
\label{alg:self-compose-prv}
\begin{algorithmic}
\STATE {\bfseries Input:} PRV $Y$, number of compositions $k = 2^t$, mesh sizes $h_1 \le \dots \le h_t$, truncation parameters $L_1 \le \dots \le L_t$, where each $L_i \in h_i \cdot (\frac{1}{2} + \bbZ_{>0})$ for all $i$.
\STATE {\bfseries Output:} PDF of an approximation $Y_{t}$ for $Y^{\oplus k}$. $Y_{t}$ will be supported on $\mu + (h_t\bbZ \cap [-L_t, L_t])$ for some $\mu \in \insquare{0, \frac{h_t}{2}}$.
\STATE $Y_0 \gets Y |_{|Y| \le L_1}$ \hfill \texttt{\small $\triangleright$ $Y$ conditioned on $|Y| \le L_1$}
\FOR{$i=0$ {\bfseries to} $t-1$}
\STATE $\wtilde{Y}_{i} \gets \DiscretizeRV(Y_{i}, h_{i+1}, L_{i+1})$
\STATE $Y_{i+1} \gets \wtilde{Y}_{i} \oplus_{L_{i+1}} \wtilde{Y}_{i}$ \hfill \texttt{\small $\triangleright$ FFT convolution}
\ENDFOR
\RETURN $Y_t$
\end{algorithmic}
\end{algorithm}

\begin{theorem}\label{thm:recursive-self-compose-error}
For all PRV $Y$ and $k = 2^t$, the approximation $Y_{t}$ returned by $\RecursiveSelfComposePRV$ satisfies
\[
    |Y^{\oplus k} - Y_{t}| \le_{\deltaerror} \epserror,
\]
using a choice of parameters satisfying for all $i \leq t$ that
\begin{align*}
h_i &\textstyle~=~ \Omega\inparen{\frac{\epserror}{t^{1.5} \sqrt{2^{t-i} \log \frac{1}{\deltaerror}}}}, \\
L_i &\textstyle~\ge~ \eps_{Y^{\oplus 2^i}}\inparen{\frac{\epserror \deltaerror}{2^{O(t)}}} + h_i \cdot \inparen{3 + 2i \sqrt{\frac{1}{2} \log \frac{2}{\eta}}}, \\
L_t & ~\ge~ \cdots ~\ge~ L_1.
\end{align*}
\end{theorem}

\paragraph{Proof Outline.} We establish coupling approximations between consecutive random variables in the sequence:
\[ Y^{\oplus 2^t},\ Y_{0}^{\oplus 2^t},\ \wtilde{Y}_{0}^{\oplus 2^t},\ Y_1^{\oplus 2^{t-1}},\ \ldots,\ Y_{t-1}^{\oplus 2},\ \wtilde{Y}_{t-1}^{\oplus 2},\ Y_t\,,\]
using a similar approach as in the proof of \cref{thm:two-stage-self-compose-error}.

\paragraph{Running time analysis.} The overall running time is at most $O\inparen{\sum_i \frac{L_i}{h_i} \log \frac{L_i}{h_i}}$, which can be upper bounded by
\begin{align*}
&\wtilde{O}\inparen{\frac{\epsupper \cdot t^{2.5} \sqrt{\log \frac{1}{\deltaerror}}}{\epserror}},
\end{align*}
where $\epsupper\textstyle~:=~ \max_{i} \inparen{\sqrt{2^{t-i}} \cdot \eps_{Y^{\oplus 2^i}}\inparen{\frac{\epserror\deltaerror}{2^{O(t)}}}}$.

In many practical regimes of interest, $\epsupper/\epserror$ is at most $\polylog(t) = \mathrm{polyloglog}(k)$. For ease of exposition in the following, we assume that $\epserror$ is a small constant, e.g. $0.1$ and suppress the dependence on $\epserror$.
Suppose the original mechanism $\calM$ underlying $Y$ satisfies $(\eps = \frac{1}{\sqrt{k \cdot \log(1/\deltaerror)}}, \delta = \frac{o_k(1)}{k^{O(1)}})$-DP.  Then by advanced composition~\cite{dwork2010boosting}, we have that $\calM^{\circ 2^i}$ satisfies $(\eps \sqrt{2^{i+1} \log \frac{1}{\delta'}} + 2^{i+1} \eps (e^{\eps} - 1), 2^i\delta + \delta')$-DP. If $2^i\delta + \delta' \lesssim \frac{\deltaerror}{2^{O(t)}}$, then we have that $\eps_{Y^{\oplus 2^i}}\inparen{\frac{\deltaerror}{2^{O(t)}}} \lesssim \sqrt{\frac{1}{2^{t-i}} \ln \frac{2^{O(t)}}{\deltaerror}}$. Instantiating this with $i = 1, \ldots, t$ gives us that $\epsupper$ is at most $\polylog(k)$.

\section{Conclusions and Discussion}
In this work, we presented an algorithm with a running time and memory usage of $\polylog(k)$ for the task of self-composing a broad class of DP mechanisms $k$ times. We also extended our algorithm to the case of composing $k$ different mechanisms in the same class, resulting in a running time and memory usage $\wtilde{O}(k)$; both of these improve the state-of-the-art by roughly a factor of $\sqrt{k}$. We also demonstrated the practical benefits of our framework compared to the state-of-the-art by evaluating on the sub-sampled Gaussian mechanism and the Laplace mechanism, both of which are widely used in the literature and in practice.

For future work, it would be interesting to tighten the log factors in our bounds. A related future direction is to make the $\RecursiveSelfComposePRV$ algorithm more practical, since the current recursive analysis is quite loose. Note that $\RecursiveSelfComposePRV$ could also be performed with an arity larger than $2$; e.g., with an arity of $100$, one would perform $100^3$ compositions as a three-stage composition. For any constant arity, our algorithm gives an asymptotic runtime of $O(\polylog k)$ as $k \to \infty$, however, for practical considerations, one may also consider adapting the arity with $k$ to tighten the log factors. We avoided doing so for simplicity, since our focus in obtaining an $O(\polylog(k))$ running time was primarily theoretical.

\section*{Acknowledgments}
We would like to thank the anonymous reviewers for their thoughtful comments that have improved the quality of the paper.

\balance
\bibliographystyle{icml2022}
\bibliography{main.bbl}

\newpage
\appendix
\onecolumn

\section{Proofs of Coupling Approximation Properties}\label{apx:gopi-proofs}
For sake of completeness, we include proofs of the lemmas we use from \citet{gopi2021numerical}.

\begin{proof}[Proof of \Cref{lem:coupling-properties}\eqref{item:coupling-triangle}]
There exists couplings $(X, Y)$ and $(Y, Z)$ such that $\Pr[|X - Y| \ge h_1] \le \eta_1$ and $\Pr[|Y - Z| \ge h_2] \le \eta_2$. From these two couplings, we can construct a coupling between $(X, Z)$: sample $X$, sample $Y$ from $Y | X$ (given by coupling $(X,Y)$) and finally sample $Z$ from $Z|Y$ (given by coupling $(Y,Z)$). Therefore for this coupling, we have:
\begin{align*}
\Pr[|X - Z| \ge h_1 + h_2] &~\le~ \Pr[|(X-Y) + (Y-Z) \ge h_1 + h_2]\\
&~\le~ \Pr[|X - Y| + |Y - Z| \ge h_1 + h_2]\\
&~\le~ \Pr[|X - Y| \ge h_1] + \Pr[|Y - Z| \ge h_2]\\
&~\le~ \eta_1 + \eta_2\,.\qedhere
\end{align*}
\end{proof}

\begin{proof}[Proof of \Cref{lem:coupling-properties}\eqref{item:coupling-composition}]
	The first part follows from the fact that there exists a coupling between $X$ and $Y$ such that $\Pr[X \ne Y] = \dTV(X, Y)$. 
	The second part follows from the first part and the fact that $\dTV(X^{\oplus k}, Y^{\oplus k}) \le k \cdot \dTV(X,Y)$.
\end{proof}

\begin{proof}[Proof of \Cref{lem:coupling-properties}\eqref{item:coupling-concentration}]
	Let $X = Y - \wtilde{Y}$ where $(Y, \wtilde{Y})$ are coupled such that $|Y - \wtilde{Y} - \mu| \le h$ with probability $1$. Then $\Ex[X] = 0$ and $X \in [\mu-h, \mu+h]$ w.p. $1$ and hence by Hoeffding's inequality,
	\[ \Pr\insquare{|X^{\oplus k}| \ge t} \le 2 \exp\inparen{-\frac{2t^2}{k(2h)^2}} = \eta\,, \qquad \text{ for } t = h \sqrt{2k \log \frac{2}{\eta}}\,. \qedhere\]
\end{proof}

\section{\boldmath Two-Stage Heterogeneous Composition}\label{apx:two-stage-extensions}

We can handle composition of $k$ different PRVs $Y^1, \ldots, Y^k$ with a slight modification to $\TwoStageSelfComposePRV$ as given in \cref{alg:two-stage-compose}.   The approximation analysis remains similar as before. The main difference is that $L_1$ and $L_2$ are to be chosen as
\begin{align*}
L_1 &~\ge~ O\inparen{\max\set{\max_i \set{\eps_{Y^i}\inparen{\frac{\epserror\deltaerror}{16k_1 k_2^{1.5}}}}\ , \ \ \max_{t} \set{\eps_{Y^{tk_1+1 : (t+1)k_1}}\inparen{\frac{\epserror\deltaerror}{64k_2^{1.5}}}}} + \frac{\epserror}{\sqrt{k_2}}},\\
L_2 &~\ge~ O\inparen{\max\set{\eps_{Y^{1:k}}\inparen{\frac{\epserror\deltaerror}{16}} + \epserror, L_1}},
\end{align*}
where we denote $Y^{i:j} = Y^i \oplus \cdots \oplus Y^j$. In the case of $k_1 = k_2 = \sqrt{k}$ (assumed to be an integer) and $r = 0$, this leads to a final running time of 
\[
\wtilde{O}\inparen{\frac{\epsupper \cdot k^{1.25} \cdot \sqrt{\log(k/\deltaerror)}}{\epserror}},
\]
where $\epsupper$ can be bounded as
\[
\max\set{
\eps_{Y^{1:k}}\inparen{\frac{\epserror\deltaerror}{16}},\ \ 
\sqrt[4]{k}\cdot \max_{t} \ \eps_{Y^{t\sqrt{k}+1 : (t+1)\sqrt{k}}}\inparen{\frac{\epserror\deltaerror}{64k^{0.75}}},\ \ 
\sqrt[4]{k}\cdot \max_i \eps_{Y^i}\inparen{\frac{\epserror\deltaerror}{16k^{1.25}}}
} \ \  + \ \ \epserror.
\]

\begin{algorithm}[t]
	\caption{$\TwoStageComposePRV$ for heterogeneous compositionss}
	\label{alg:two-stage-compose}
	\begin{algorithmic}
		\STATE {\bfseries Input:} PRVs $Y^1, \ldots, Y^k$, number of compositions $k = k_1 \cdot k_2 + r$, $r < k_1$, mesh sizes $h_1 \le h_2$, truncation parameters $L_1 \le L_2$, where each $L_i \in h_i \cdot (\frac{1}{2} + \bbZ_{>0})$.
		\STATE {\bfseries Output:} PDF of an approximation $\what{Y}_{2}$ for $Y^1 \oplus \ldots \oplus Y^k$. $\what{Y}_{2}$ will be supported on $\mu + (h_2\bbZ \cap [-L_2, L_2])$ for some $\mu \in \insquare{-\frac{h_2}{2}, \frac{h_2}{2}}$.
		\STATE
		\FOR{$i = 1$ to $k_1k_2$}
    		\STATE $Y_0^i \gets Y^i |_{|Y^i| \le L_1}$ \hfill \texttt{ \small $\triangleright$ $Y^i$ conditioned on $|Y^i| \le L_1$}
    		\STATE $\wtilde{Y}^i_{0} \gets \DiscretizeRV(Y^i_{0}, h_{1}, L_{1})$
		\ENDFOR
		\FOR{$i = k_1k_2+1$ to $k$}
		\STATE $Y_0^i \gets Y^i |_{|Y^i| \le L_2}$ \hfill \texttt{ \small $\triangleright$ $Y^i$ conditioned on $|Y^i| \le L_2$}
		\STATE $\wtilde{Y}^i_{0} \gets \DiscretizeRV(Y^i_{0}, h_{2}, L_{2})$
		\ENDFOR
		\STATE
		\FOR{$t = 0$ to $k_2-1$}
		    \STATE $Y^t_1 \gets \wtilde{Y}^{tk_1 + 1}_0 \oplus_{L_1} \cdots \oplus_{L_1} \wtilde{Y}^{(t+1)k_1}_0$ \hfill \texttt{\small $\triangleright$ $k_1$-fold FFT convolution}
		    \STATE $\wtilde{Y}^t_{1} \gets \DiscretizeRV(Y^t_{1}, h_{2}, L_{2})$
	    \ENDFOR
	    \STATE
		\STATE $Y_{2} \gets \wtilde{Y}^1_{1} \oplus_{L_2} \cdots \oplus_{L_2} \wtilde{Y}^{k_2}_1$ \hfill \texttt{\small $\triangleright$ $k_2$-fold FFT convolution}
		\RETURN $Y_2 \oplus_{L_2} 
		\wtilde{Y}_0^{k_1 \cdot k_2 + 1} \oplus_{L_2} \cdots \oplus_{L_2}
		\wtilde{Y}_0^{k}$
	\end{algorithmic}
\end{algorithm}

\section{\boldmath Analysis of $\RecursiveSelfComposePRV$}

We establish coupling approximations between consecutive random variables in the sequence:
\[ Y^{\oplus 2^t},\ Y_{0}^{\oplus 2^t},\ \wtilde{Y}_{0}^{\oplus 2^t},\ Y_1^{\oplus 2^{t-1}},\ \wtilde{Y}_1^{\oplus 2^{t-1}}, \ Y_2^{\oplus 2^{t-2}},\ \ldots,\ Y_{t-1}^{\oplus 2},\ \wtilde{Y}_{t-1}^{\oplus 2},\ Y_t\,, \]

\paragraph{\boldmath Coupling $Y^{\oplus 2^t}$ and $Y_0^{\oplus 2^t}$.}
Since $\dTV(Y, Y_0) = \Pr[|Y| > L_1] =: \delta_0$, we have from \cref{lem:coupling-properties}\eqref{item:coupling-composition} that
\begin{align}
	|Y^{\oplus 2^t} - Y_0^{\oplus 2^t}| &~\le_{2^t \delta_0}~ 0\,.\label{eq:base-case}
\end{align}

\paragraph{\boldmath Coupling $Y_i^{\oplus 2^{t-i}}$ and $\wtilde{Y}_{i}^{\oplus 2^{t-i}}$.}
We have from \cref{claim:discretize-prv-prop} that $\Ex[Y_i] = \Ex[\wtilde{Y}_i]$ and that $|Y_i - \wtilde{Y}_i - \mu_i| \le_0 \frac{h_{i+1}}{2}$ for some $\mu_i$ satisfying $|\mu_i| \le \frac{h_{i+1}}{2}$. Thus, applying \cref{lem:coupling-properties}\eqref{item:coupling-concentration}, we have (for $\eta$ to be chosen later)  that
\begin{align}
	\inabs{Y_i^{\oplus 2^{t-i}} - \wtilde{Y}_{i}^{\oplus 2^{t-i}}} ~\le_{\eta}~ h_{i+1} \sqrt{2^{t-i-1} \log \frac{2}{\eta}} ~=:~ \wtilde{h}_{i+1}. \label{eq:re-discretize}
\end{align}

\paragraph{\boldmath Coupling $\wtilde{Y}_i^{\oplus 2^{t-i}}$ and $Y_{i+1}^{\oplus 2^{t-i-1}}$.}
Since $\dTV(\wtilde{Y}_i \oplus \wtilde{Y}_i, \wtilde{Y}_i \oplus_{L_{i+1}} \wtilde{Y}_i) \le \Pr[|\wtilde{Y}_i \oplus \wtilde{Y}_i| > L_{i+1}] =: \delta_{i+1}$, it holds via \cref{lem:coupling-properties}\eqref{item:coupling-composition} that
\begin{align}
	|\wtilde{Y}_i^{\oplus 2^{t-i}} - Y_{i+1}^{\oplus 2^{t-i-1}}| \le_{2^{t-i-1} \delta_{i+1}} 0\,.\label{eq:wrap-around}
\end{align}

\paragraph{Putting things together.}
Thus, combining \cref{eq:re-discretize,eq:wrap-around} for $i \in \set{0, \ldots, t-1}$, using \cref{lem:coupling-properties}\eqref{item:coupling-composition}, we have that
\begin{align}
	& \hspace{-1cm}\inabs{Y_0^{\oplus 2^t} - Y_t} ~\le_{\delta^*}~ \eps^* \label{eq:Y_t-approx}, \\
	\text{where } \qquad \delta^* ~:=~ \delta_t^* &\textstyle~:=~ t\eta + \sum_{j=1}^t 2^{t-j} \delta_j\,, \label{eq:delta-star}\\
	\text{and } \qquad \eps^* ~:=~ \eps_t^* &\textstyle~:=~ \sum_{j=1}^t h_j \sqrt{2^{t-j} \log \frac{2}{\eta}} \nonumber.
\end{align}
More generally, the same analysis shows that for any $1 \le i \le t$,
\begin{align}
& \inabs{Y_0^{\oplus 2^i} - Y_i} ~\le_{\delta_i^*}~ \eps_i^*, \nonumber\\
\text{where } \qquad \delta_i^* &\textstyle~:=~ i\eta + \sum_{j=1}^i 2^{i-j} \delta_j\,, \label{eq:delta-i-star}\\
\text{and } \qquad \eps_i^* &\textstyle~:=~ \sum_{j=1}^i h_j \sqrt{2^{i-j} \log \frac{2}{\eta}}\,. \nonumber
\end{align}
To simplify our analysis going forward, we fix the choice of $h_i$'s that we will use, namely, $h_i = \frac{\eps^*}{t \sqrt{2^{t-i} \log \frac{2}{\eta}}}$ (for $\eta$ that will be chosen later).
This implies that
\begin{align*}
\eps_i^* &~=~ \sum_{j=1}^i h_j \sqrt{2^{i-j} \log \frac{2}{\eta}} ~=~ i \cdot \frac{\eps^*}{t \sqrt{2^{t-i}}}
~=~ h_{i+1} \cdot i \sqrt{\frac{1}{2}\log\frac{2}{\eta}}, 
\end{align*}
where the last step uses that $i \le t$.%

In order to get our final bound, we need to bound $\delta_i$'s in terms of $\eta$, the mesh sizes $h_i$'s, and truncation parameters $L_i$'s. For ease of notation, we let $\delta_0^* = 0$. We have for $0 \le i < t$ that 
\begin{align}
\delta_{i+1} &~=~ \Pr\insquare{\inabs{\wtilde{Y}_i \oplus \wtilde{Y}_i} > L_{i+1}} \nonumber\\
&~\le~ \Pr\insquare{\inabs{Y_i \oplus Y_i} > L_{i+1} - 2h_{i+1}} \qquad \text{(since, $\inabs{Y_i - \wtilde{Y}_i} \le h_{i+1}$ w.p. $1$)} \nonumber\\
&~\le~ 2 \Pr\insquare{\inabs{Y_i - Y_0^{\oplus 2^i}} > \eps_i^*} + \Pr\insquare{\inabs{Y_0^{\oplus 2^{i+1}}} > L_{i+1} - 2h_{i+1} - 2 \eps_{i}^*} \nonumber\\
&~\le~ 2 \delta_i^* + \Pr\insquare{\inabs{Y^{\oplus 2^{i+1}}} > L_{i+1} - 2h_{i+1} - 2 \eps_{i}^*}\nonumber\\
\delta_{i+1} &~\le~ 2 \delta_i^* + \frac{4}{\alpha_{i+1}}\delta_{Y^{\oplus 2^{i+1}}}(\wtilde{L}_{i+1})\label{eq:delta-from-delta-star},
\end{align}
where in the penultimate step we use that the tails of $Y_0^{\oplus 2^i}$ are no larger than tails of $Y^{\oplus 2^i}$ since $Y_0$ is a truncation of $Y$, and in the last step we use \cref{lem:tail-bounds} with $\wtilde{L}_{i+1} := L_{i+1} - 2 h_{i+1} (1 + i \sqrt{\frac{1}{2}\log\frac{2}{\eta}}) - \alpha_{i+1}$ (eventually we set $\alpha_{i+1} = h_{i+1}$).

We show using an inductive argument that for $C_i = 8^i$,
\begin{align}
\delta_i &\textstyle~\le~ 2C_i \cdot \inparen{\eta + \sum_{j=1}^i \frac{4}{\alpha_{j}} \delta_{Y^{\oplus 2^j}}\inparen{\wtilde{L}_j}}\,,\label{eq:inductive-step-1}\\
\text{and}\qquad \delta_i^* &\textstyle~\le~ C_{i+1} \cdot \inparen{\eta + \sum_{j=1}^i \frac{4}{\alpha_{j}} \delta_{Y^{\oplus 2^j}}\inparen{\wtilde{L}_j}}. \label{eq:inductive-step-2}
\end{align}
The base case holds since $\delta_1 ~\le~ \frac{4}{\alpha_1} \delta_{Y^{\oplus 2}}(\wtilde{L}_1)$; note $C_1 > 1$. From \eqref{eq:delta-i-star}, we have
\begin{align*}
    \delta_i^* &\textstyle~\le~ i\eta + \sum_{j=1}^i 2^{i-j} \delta_j \\
    &\textstyle~\le~ i\eta + \sum_{j=1}^i 2^{i-j}  \left(2C_j \cdot \inparen{\eta + \sum_{\ell=1}^j \frac{4}{\alpha_{\ell}} \delta_{Y^{\oplus 2^\ell}}\inparen{\wtilde{L}_\ell}}\right) \\
    &\textstyle~\le~ i\eta + \inparen{\sum_{j=1}^i 2^{i-j} \cdot 2C_j} \cdot \inparen{\eta + \sum_{j=1}^i \frac{4}{\alpha_{j}} \delta_{Y^{\oplus 2^j}}\inparen{\wtilde{L}_j}} \\
    &\textstyle~\le~ i\eta + (4C_i) \cdot \inparen{\eta + \sum_{j=1}^i \frac{4}{\alpha_{j}} \delta_{Y^{\oplus 2^j}}\inparen{\wtilde{L}_j}} \\
    &\textstyle~\le~ C_{i+1} \cdot \inparen{\eta + \sum_{j=1}^i \frac{4}{\alpha_{j}} \delta_{Y^{\oplus 2^j}}\inparen{\wtilde{L}_j}}.
\end{align*}
This completes the inductive step \eqref{eq:inductive-step-2}. Finally, \eqref{eq:delta-from-delta-star} immediately implies the inductive step \eqref{eq:inductive-step-1}.

Putting this together in \eqref{eq:delta-star}, and setting $\alpha_{i} = h_i$, we get
\begin{align*}
	\delta_t^* &~\le~ \frac{1}{\eps^*} \cdot 2^{O(t)} \inparen{\eta + \sum_{i=1}^t \delta_{Y^{\oplus i}}(\wtilde{L}_i)}. %
\end{align*}
Finally, combining \eqref{eq:Y_t-approx} with \eqref{eq:base-case} using \cref{lem:coupling-properties}\eqref{item:coupling-triangle}, we get
\begin{align}
	& \inabs{Y^{\oplus 2^t} - Y_t} ~\le_{\deltaerror}~ \epserror \nonumber \\
	\text{where } \qquad \deltaerror &\textstyle~:=~ \frac{1}{\eps^*} \cdot \inparen{2^{O(t)} \inparen{\eta + \sum_{i=1}^t \delta_{Y^{\oplus 2^i}}(\wtilde{L}_{i})} + 2^{O(t)} \delta_Y(\wtilde{L}_1)}\,, \nonumber \\
	\text{and } \qquad \epserror &\textstyle~:=~ \eps^*.
	\nonumber
\end{align}
Thus, we get the desired approximation result for the following choice of parameters
\begin{align*}
    \eta &\textstyle~=~ \frac{\deltaerror}{2^{O(t)}}, \\
    h_i &\textstyle~=~ \frac{\epserror}{t \sqrt{2^{t-i} \log \frac{2}{\eta}}} ~=~ \Omega\inparen{\frac{\epserror}{t^{1.5} \sqrt{2^{t-i} \log \frac{1}{\deltaerror}}}}, \\
    L_i &\textstyle~\ge~ \max\set{\eps_{Y^{\oplus 2^i}}\inparen{\frac{\epserror \deltaerror}{2^{O(t)}}} + h_i \cdot \inparen{3 + 2i \sqrt{\frac{1}{2} \log \frac{2}{\eta}}}, L_{i-1}}.
\end{align*}

Thus, the overall running time is at most
\begin{align*}
\wtilde{O}\inparen{\sum_i \frac{L_i}{h_i}} &~=~ \wtilde{O}\inparen{\frac{\epsupper \cdot t^{2.5} \sqrt{\log \frac{1}{\deltaerror}}}{\epserror}}, \\
\text{where } \qquad \epsupper&\textstyle~:=~ \max_{i} \inparen{\sqrt{2^{t-i}} \cdot \eps_{Y^{\oplus 2^i}}\inparen{\frac{\epserror\deltaerror}{2^{O(t)}}}}.
\end{align*}

\paragraph{\boldmath Extensions of $\RecursiveSelfComposePRV$.} Similar to \cref{apx:two-stage-extensions}, it is possible to extend $\RecursiveSelfComposePRV$ to handle the case where the number of compositions $k$ is not a power of $2$, with a similar asymptotic runtime. It can then be extended to handle heterogeneous compositions of $k$ different mechanisms.
\end{document}